\documentclass[11pt,a4paper,oneside]{article}
\usepackage{psfig,times,natbib}
\def\et{{et al.\ }}
\def\mcg{{MCG--6-30-15}}
\def\sax{{\it BeppoSAX}}
\def\xte{{\it RXTE}}
\def\asca{{\it ASCA}}
\def\xmm{{\it XMM-Newton}}

\def\axaf{{\it Chandra}}

\def\km{{\rm\thinspace km}}

\def\Msun{\hbox{$\rm\thinspace M_{\odot}$}}

\def\s{{\rm\thinspace s}}

\def\kmps{\hbox{$\km\s^{-1}\,$}}

\author{Andrew C. Fabian and Giovanni Miniutti \\
  {\small{ Institute of Astronomy, Madingley Road Cambridge CB3 0HA, UK}}}

\title{The X-ray spectra of accreting Kerr black holes}

\begin{document}
\maketitle
%\clearpage
%\tableofcontents

\begin{abstract} 
\noindent The relativistic broad iron lines seen in the X-ray
spectra of several active galaxies and Galactic black hole systems are
reviewed. Most such objects require emission from within the innermost
stable orbit of a non-rotating black hole, suggesting that the black
holes are rapidly spinning Kerr holes. We discuss the soft excess, the
broad iron line and the Compton hump characteristic of reflection from
partially ionized gas and show that they may be a common ingredient in
the X-ray spectra of many radiatively-efficient, accreting black
holes. Strong gravitational bending of the radiation close to a Kerr
black hole can explain the otherwise puzzling spectral variability
seen in some objects. The Narrow Line Seyfert 1 galaxies may be among
the most extreme objects yet seen.

\end{abstract}

\clearpage

%\chapter{The X-ray spectra of Kerr black holes}

\section{Introduction}

The form of the spacetime geometry around an astrophysical black hole
(BH) is due to Roy Kerr (1963) who found the exact solution to the
general relativistic Einstein's equations for the spacetime outside
the horizon of a rotating BH.  Since then, rotating BH are known as
Kerr BH, the spacetime geometry of which depends on two parameters only, the
BH mass and spin. In the limit of no rotation, Kerr's solution
coincides with that for a non--rotating BH found earlier by Karl
Schwarzschild (1916), which obviously depends on the BH mass only. The
complete solution of the equations of motion in the Kerr spacetime is
due to Brandon Carter (1968). The interested reader finds an excellent
treatment of Einstein's theory of General Relativity and of BH
spacetimes in the fundamental book {\it Gravitation} by Misner, Thorne
\& Wheeler (1973).

A large fraction of the accretion energy in luminous BH systems is
dissipated in the innermost regions of the accretion
flow. Most of the power is radiated from close to the smallest
accretion disc radii in the relativistic region close to the BH
(Shakura \& Sunyaev 1973; Pringle 1981).  Relativistic effects then
affect the appearance of the X--ray spectrum from the disc through
Doppler, aberration, gravitational redshift and light bending effects
(Page \& Thorne 1974; Cunningham 1975, 1976). The dominant feature in
the 2--10~keV X--ray spectrum seen by a distant observer is an iron line with a
broad skewed profile which carries unique information on the
structure, geometry, and dynamics of the accretion flow in the
immediate vicinity of the central BH, providing a tool to
investigate the nature of the spacetime there (Fabian et al.\ 1989;
Laor 1991).

Relativistic broad iron lines seen in the spectrum of several active
galaxies and Galactic black hole binaries are reviewed here (see also
Fabian et al 2000; Reynolds \& Nowak 2003 for previous reviews). Among
others, the cases for relativistic lines in the Seyfert galaxies
MCG--6-30-15 and IRAS~18325--5926, and the X-ray binaries
XTE~J1650--500 and GX\,339-4 are very strong (Tanaka et al 1995; Wilms
et al 2001; Fabian et al 2002a; Iwasawa et al 1996a; Iwasawa et al
2004; Miller et al 2002a; Miniutti, Fabian \& Miller 2004; Miller et
al 2004a,b). In three out of four objects, the X--ray data require
emission from within the innermost stable circular orbit of a
non--rotating BH, suggesting that the BHs in many objects are rapidly
spinning.  The spectra of many other objects, in particular Narrow
Line Seyfert 1 (NLS1) galaxies, can be successfully described by a
simple two--component model comprising a highly variable power law
continuum and a much more constant reflection component from the
accretion disc (e.g.\ Fabian et al 2004; Fabian et al 2005; Ponti et al
2005).  The puzzling spectral variability of such sources is now
beginning to be understood within the context of emission from the
strong gravity regime (Miniutti et al 2003; Miniutti \& Fabian 2004).
Some active galactic nuclei (AGN) and X-ray black hole binaries show
either no line or only a narrow one (e.g.\ Page et al 2004; Yaqoob \&
Padmanabhan 2004).  This is discussed within the context of both
theoretical implications and present observational limitations.
Finally, the short--timescale variability of the broad iron line is
beginning to be unveiled, providing exciting results that are
dramatically improving our understanding of the innermost regions of
the accretion flow, where General Relativity is no longer a small
correction and becomes the most relevant physical ingredient (e.g.\
Turner et al 2002; Iwasawa, Miniutti \& Fabian 2004).

The advent of future X--ray missions with higher energy resolution
(e.g. {\it Suzaku} succesfully launched in July 2005) and much larger
collecting area in the relevant iron band (XEUS and Constellation--X
in the next decade) will open up a new window on the innermost regions
of the accretion flow in AGN and X--ray binaries. This will enable us
to further test the general picture we propose here, with the
potential of mapping with great accuracy the strong field regime of
General Relativity in a manner which is still inaccessible at other
wavelengths.

\section{Main components of the X--ray spectrum}

\begin{figure}[t]
\begin{center} 
\hbox{
\psfig{figure=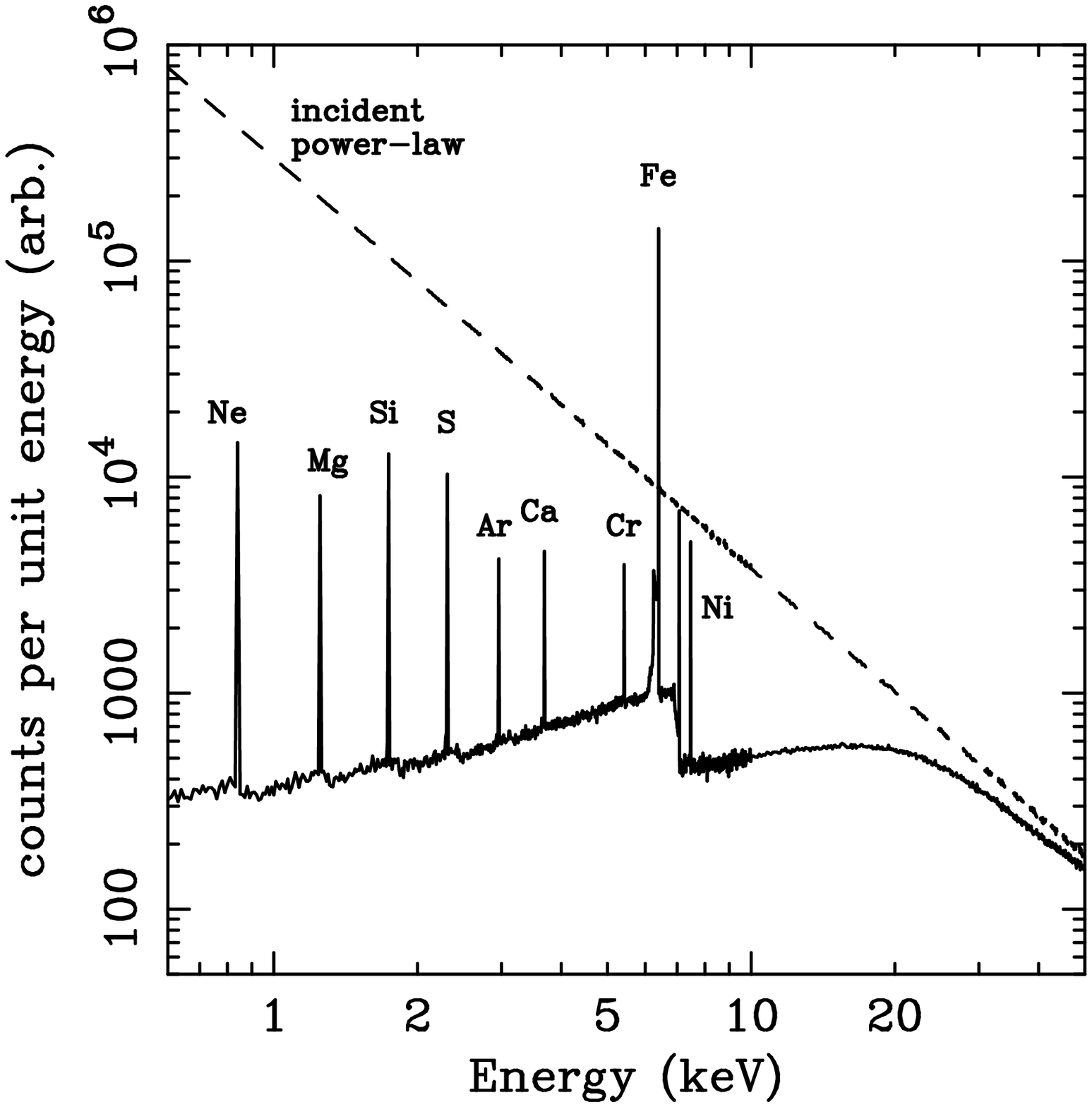,width=0.65\textwidth,height=0.36\textwidth,angle=-90}
\hspace{-2.5cm}
\psfig{figure=xrayspec1.ps,width=0.45\textwidth,height=0.32\textwidth,angle=-90}
}
\vspace{-1cm} 
\end{center} 
\caption{{\footnotesize{ {\it Left panel}: Monte Carlo simulations of
      the reflection spectrum from a slab of uniform density neutral
      matter with solar abundances. The incident power law continuum
      is also shown. Figure from Reynolds (1996). {\it Right panel}: The
      main components of the X--ray spectra of unobscured accreting BH
      are shown: soft quasi--thermal X--ray emission from the
      accretion disc (red); power law from Comptonization of the soft
      X--rays in a corona above the disc (green); reflection continuum
      and narrow Fe line due to reflection of the hard X--ray emission
      from dense gas (blue).  }}}
\label{xrayspec} 
\end{figure}

Radiatively-efficient accreting BHs are expected to be surrounded by a
dense disc radiating quasi-blackbody thermal EUV and soft X-ray
emission (see e.g the seminal paper by Shakura \& Sunyaev 1973).
However, the X--ray spectra of accreting BHs also exhibit a power law
component extending to hard X--ray energies up to 200~keV or more. The
most promising physical mechanism to produce such hard power law
components is Comptonization of the soft X--ray photons in a corona
above the disc, possibly fed by magnetic fields from the body of the
disc itself (e.g.\ Haardt \& Marschi 1991 and 1993, Zdziarski et al
1994). Irradiation of the dense disc material by the hard X-rays then
gives rise to a characteristic ``reflection'' spectrum which is the
result of Compton scattering and photoelectric absorption followed
either by Auger de-excitation or by fluorescent line emission (see
e.g.\ Guilbert \& Rees 1988; Lightman \& white 1988; George \& Fabian
1991, Matt, Perola \& Piro 1991).  This last process gives rise to an
emission line spectrum where fluorescent narrow K$\alpha$ lines from
the most abundant metals are seen. Thanks to a combination of large
cosmic abundance and high fluorescent yield, the iron (Fe) K$\alpha$
line at 6.4~keV is the most prominent fluorescent line in the X--ray
reflection spectrum. Photoelectric absorption is an
energy--dependent process, so that incident soft X--rays are mostly
absorbed, whereas hard photons tend to be Compton scattered back out
of the disc.  However, above a few tens of keV, Compton recoil reduces
the backscattered photon flux and produces a broad hump--like
structure around 20--30~keV (the so--called Compton hump). An example
of the X--ray reflection spectrum from a neutral and uniform density
semi--infinite slab of gas is shown in the left panel of
Fig.~\ref{xrayspec} where fluorescent emission lines dominate below
about 8~keV and the Compton hump is seen above 20~keV (from Reynolds
1996).

In the case of AGNs, the accretion disc is not the only reflector able
to produce a X--ray reflection spectrum.  The presence of a dusty
molecular torus surrounding the accreting system is required by
unification models at the parsec scale (Antonucci 1993). The torus
provides the absorbing column (sometimes more than
$10^{24}$~cm$^{-2}$) that is observed in Seyfert 2 galaxies which are
thought to be distinguished from unobscured Seyfert 1 galaxies only
because of a higher observer inclination. The torus itself is
Compton--thick and therefore provides an additional reflector
producing a spectrum very similar to that seen in the left panel of
Fig.~\ref{xrayspec}. As will be explained in detail in the next
section, special and general relativistic effects shape the reflection
spectrum from the centre of the accretion disc and not that from
regions far away from the central BH (such as the torus), helping us
to disentangle the two contributions.

In the right panel of Fig~\ref{xrayspec}, we show the main components
of the X--ray spectrum of accreting black holes. The ``soft excess''
represents here the soft X--ray emission from the disc, although some
other interpretations are possible and will be discussed throughout
this contribution. We also show the power law (with a high energy
cut--off characteristic of the coronal temperature) and reflection
components. Reflection is here represented by the reflection continuum
(characterised by the ``Compton hump'') and the Fe K$\alpha$ line at
6.4~keV that we assume here to be narrow, i.e.\ emitted from a distant
reflector such as the torus. The spectrum is absorbed by a column of
$2\times 10^{20}$~cm$^{-2}$, representing the typical value for
absorption by our Galaxy in the line of sight.
\begin{figure}[]
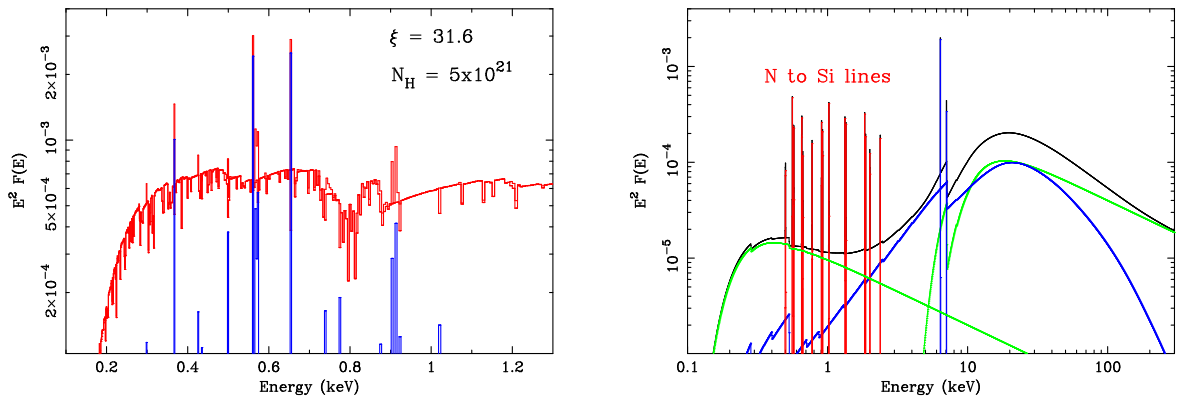

\begin{center} 
\hbox{
\psfig{figure=warmabs1.ps,width=0.45\textwidth,height=0.32\textwidth,angle=-90}
\hspace{0.8cm}
\psfig{figure=sey2.ps,width=0.45\textwidth,height=0.32\textwidth,angle=-90}
}
\vspace{-1cm} 
\end{center} 
\caption{{\footnotesize{ {\it Left panel}: The effect of photoionized
      gas surrounding an AGN is shown both in absorption (red) and
      emission (blue).  The spectrum is from the {\tt XSTAR} code of
      Kallman \& Krolik (1986).  {\it Right panel}: The typical
      spectrum of a Seyfert 2 galaxy.  The primary continuum (green)
      is heavily absorbed by a large column (typically larger than
      $10^{24}$~cm$^{-2}$) which is identified as the torus,
      partially blocking the line of sight. The same column produces
      nearly neutral X--ray reflection (blue) from the visible side of
      the torus. Emission lines due to photoionized gas (red) are
      detectable because of the reduced primary continuum flux.  }}}
\label{wabs} 
\end{figure}

Another important component in the X--ray spectrum of AGN is due to
the ubiquitous presence of warm gas surrounding the central nucleus.
This gas is photoionized by the primary X--ray continuum and
contributes to the spectrum both in absorption and in re--emission
(Halpern 1984; Reynolds 1997; Kaastra et al 2000; Kaspi et al 2001).
In the case of Seyfert 1 galaxies, the continuum level is so high that
the emission component is diluted and generally not observable, except
in particularly low flux states of the X--ray source. One remarkable
case in this sense is provided by the low flux state of NGC~4051 (see
the analysis of the high resolution \xmm--RGS data below 2~keV by
Pounds et al 2004b; also Ponti et al 2005). 

In Seyfert 2 galaxies, the primary continuum is heavily absorbed by
large columns in the line of sight (the torus) and therefore the
emission component is readily detectable. On the other hand,
absorption by this same photoionized gas is obviously easier to detect
in Seyfert 1 than Seyfert 2 galaxies due to the higher level of the
continuum. In the left panel of Fig.~\ref{wabs} we show both
absorption and emission components on a primary power law continuum
(plus cold absorption from our own Galaxy in the direction of the
source) in the most relevant soft X--ray band. The gas is assumed to
have a column of $N_{\rm{H}} = 5\times 10^{21}$~cm$^{-2}$ and
ionization parameter of $\xi = 31.6$~erg~cm~s$^{-1}$ and the spectrum
is modelled with the {\tt XSTAR} code (Kallman \& Krolik 1986). In the
right panel of the same figure, we show a qualitative view of the
typical spectrum for a Seyfert 2 galaxy. The hard spectrum is
dominated by reflection from the torus, whereas in the soft band
emission lines from photoionized gas can be detected because of the
low continuum level (e.g.\ Kinkhabwala et al 2002; Bianchi et al 2005
and many others).  In the following we shall focus on unobscured
accretion systems such as Sefert 1 galaxies and BH binaries in which
the innermost regions of the accretion flow are in principle
observable.

\section{The relativistic iron line}  

The Fe\,I K$\alpha$ emission line is a doublet comprising K$\alpha 1$
and K$\alpha 2$ lines at 6.404~keV and 6.391~keV respectively.
However, the energy separation has generally been too small to be
detected and a weighted mean at 6.4~keV is generally assumed. The
resulting line is symmetric and intrinsically much narrower than the
spectral resolution commonly available today\footnote{{\it Suzaku} will soon
provide the high energy resolution needed to resolve the K$\alpha 1$
and K$\alpha 2$ lines}. Hence, the detailed line shape and broadening
can be used to study the dynamics of the emitting region.

\begin{figure}[t!]
\hspace{-1cm}
\psfig{figure=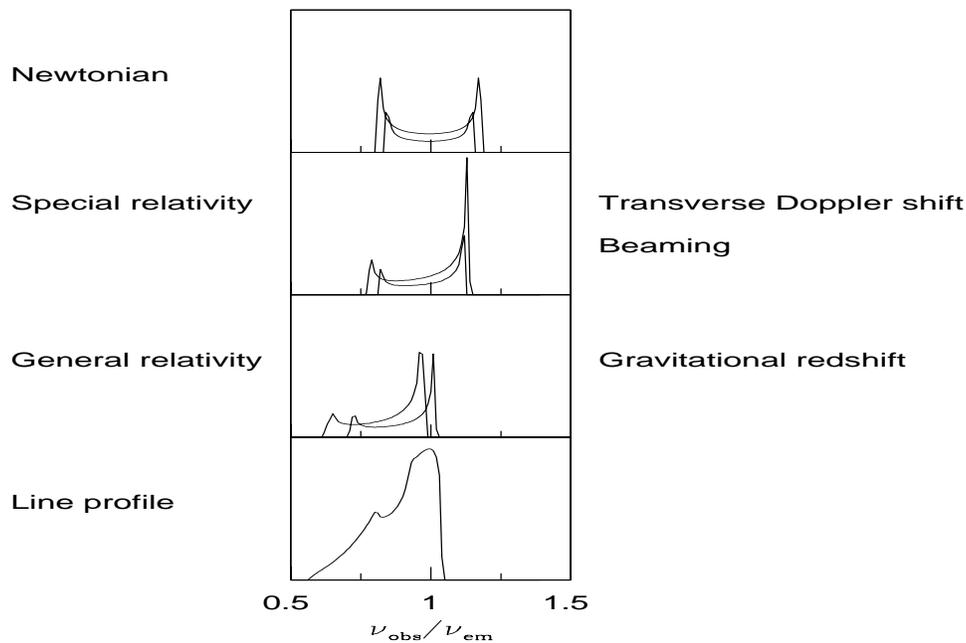,width=1.2\textwidth,height=0.7\textwidth}
\vspace{-2.8cm}
\caption{{\footnotesize{The profile of an intrinsically narrow
      emission line is modified by the interplay of
      Doppler/gravitational energy shifts, relativistic beaming, and
      gravitational light bending occurring in the accretion disc
      (from Fabian et al 2000). The upper panel shows the symmetric
      double--peaked profile from two annuli on a non--relativistic
      Newtonian disc. In the second panel, the effects of transverse
      Doppler shifts (making the profiles redder) and of relativistic
      beaming (enhancing the blue peak with respect to the red) are
      included. In the third panel, gravitational redshift is turned
      on, shifting the overall profile to the red side and reducing
      the blue peak strength. The disc inclination fixes the maximum
      energy at which the line can still be seen, mainly because of
      the angular dependence of relativistic beaming and of
      gravitational light bending effects.  All these effects combined
      give rise to a broad, skewed line profile which is shown in the
      last panel, after integrating over the contributions from all
      the different annuli on the accretion disc.}}}
\label{profiles} 
\end{figure}

If the reflection spectrum, and therefore the Fe line, originates from
the accretion disc, the line shape is distorted by Newtonian, special
and general relativistic effects (see e.g.\ Fabian et al 2000). This is
illustrated schematically in Fig.~\ref{profiles}. In the Newtonian
case, each radius on the disc produces a symmetric double--peaked line
profile with the peaks corresponding to emission from the approaching
(blue) and receding (red) sides of the disc. Close to the BH, where
orbital velocities become relativistic, relativistic beaming enhances
the blue peak with respect to the red one, and the transverse Doppler
effect shifts the profile to lower energies. As we approach the
central BH and gravity becomes strong enough, gravitational redshift
becomes important with the effect that the overall line profile is
shifted to lower energies.  The disc inclination fixes the maximum
energy at which the line can still be seen, mainly because of the
angular dependence of relativistic beaming and of gravitational light
bending effects.  Integrating over all radii on the accretion disc, a
broad and skewed line profile is produced, such as that shown in the
bottom panel of Fig.~\ref{profiles}. It is clear from the above
discussion that the detailed profile of a broad relativistic line from
the accretion disc has the extraordinary potential of revealing the
dynamics of the innermost accretion flow in accreting BHs and even of
testing Einstein's theory of General Relativity in a manner that is
unaccessible to other wavelengths.

\subsection{Dependence on disc inclination and emissivity}

\begin{figure}[t]
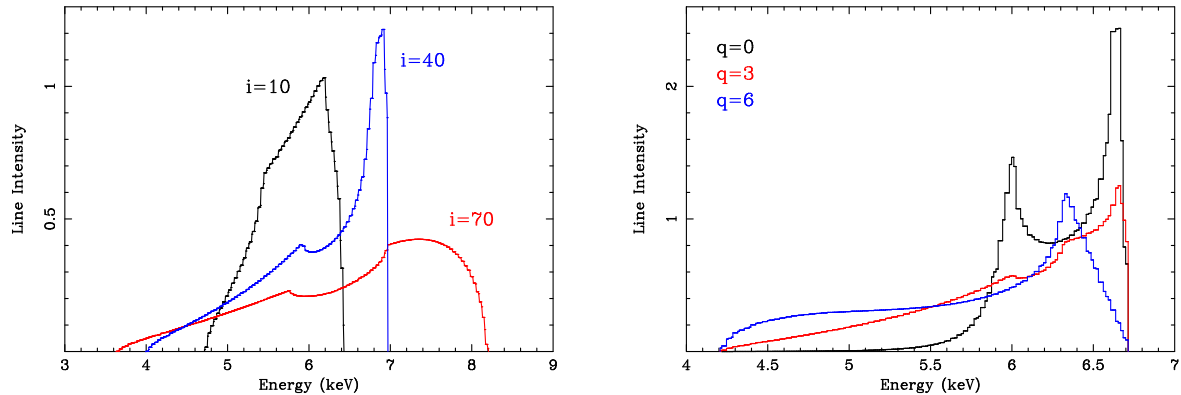
 
\begin{center} 
\hbox{
\psfig{figure=lineinc.ps,width=0.45\textwidth,height=0.32\textwidth,angle=-90}
\hspace{0.8cm}
\psfig{figure=lineq.ps,width=0.45\textwidth,height=0.32\textwidth,angle=-90}
}
\vspace{-1cm} 
\end{center} 
\caption{{\footnotesize{ {\it Left panel}: The dependence of the line
      profile on the observer inclination is shown. {\it Right
        panel}: The dependence of the line profile on the emissivity
      profile on the disc is shown. The disc emissivity is assumed to
      scale as $\epsilon = r^{-q}$. The steeper the emissivity, the
      broader and more redshifted the line profile, because more
      emphasis is given to the innermost radii where gravity
      dominates.  }}}
\label{linedep}
\end{figure}

The relativistic line profile exhibits a dependence for many physical
parameters. The energy of the blue peak of the line is mainly dictated
by the inclination of the observer line of sight with respect to the
accretion disc axis. This is clear in the left panel of
Fig.~\ref{linedep} where we show the result of fully relativistic
computations (e.g.\ Fabian et al 1989; Laor 1991; Dov\v{c}iak, Karas \&
Yaqoob 2004 among many others). The three profiles  have all the
same parameters but different observer inclination $i$.  From the Figure, it
is clear that the higher the inclination the bluer the line is,
providing a quite robust tool to measure the inclination of the
accretion disc. 

Another important parameter is the form of the emissivity profile,
i.e.\ the efficiency with which the line is emitted as a function of
the radial position on the disc. This depends mainly on the
illumination profile by the hard X--rays from the corona which is in
turn determined by the energy dissipation on the disc and by the
heating events in the corona (possibly associated with magnetic fields
see e.g.\ Merloni \& Fabian 2001a,b).  The emissivity profile is
generally assumed to be in the form of a simple power law
$\epsilon(r)=r^{-q}$, where $q$ is the emissivity index (but see e.g.\
Beckwith \& Done 2004). By assuming that the emissivity is a good
tracer of the energy dissipation on the disc, the standard value for
the emissivity index is $q=3$ (e.g.\  Pringle 1981; also Reynolds \&
Nowak 2003 and Merloni \& Fabian 2003 for a discussion on the
dependence of the emissivity profile on boundary conditions). In the
right panel of Fig.~\ref{linedep}, we show the dependence of the line
profile from this most important parameter. We show the cases of a
uniform ($q=0$), standard ($q=3$), and steep ($q=6$) emissivity
profile. A steep emissivity profile indicates that the conversion of
the X--ray photons from soft to hard in the corona is centrally
concentrated thereby illuminating more efficiently the very inner
regions of the accretion disc. As shown in the figure, steeper
emissivity profiles produce much broader and redshifted lines because
more weight is given to the innermost disc, where gravitational
redshift dominates.

\subsection{Self--consistent ionized reflection models}

So far, we have assumed that the disc (or to be more precise its outer
layers) is a slab of uniform density gas where hydrogen and helium are
fully ionized, but all the metals are neutral. The real situation is
likely to be much more complex. One first important step towards the
accurate model of accretion disc atmospheres is made by considering
thermal and ionization equilibrium.  Results of such computations have
been published over the last ten years or so with different degrees of
complexity (e.g.\ Ross \& Fabian 1993; Matt, Fabian \& Ross 1993,
1996; Zycki et al 1994; Nayakshin, Kazanas \& Kallman 2000;
R\'oza\'nska et al 2002; Dumont et al 2003).  See Ballantyne, Ross \&
Fabian 2001 for a comparison between different hypothesis such as
constant--density atmospheres and atmospheres in hydrostatic
equilibrium. The recent work by Ross \& Fabian (2005) extending and
improving previous computations (e.g.\ Ross, Fabian \& Young 1999;
Ballantyne, Ross \& Fabian 2001) is described here in some detail
since it is used extensively in comparing X--ray data to theoretical
models.

The illuminating radiation is assumed to have an exponential cut--off
power law form with high--energy cut--off fixed at 300~keV and
variable photon index $\Gamma$ between 1 and 3 roughly covering the
observed range. The ionization parameter $\xi$ is defined as the ratio
between the isotropic total illuminating flux and the comoving
hydrogen number density of the gas; results are produced for $\xi$
ranging from $1$~erg~cm~s$^{-1}$ to $10^4$~erg~cm~s$^{-1}$.  The local
temperature and fractional ionization of the gas are computed
self--consistently by solving the equations of thermal and ionization
equilibrium and ions from C, N, O, Ne, Mg, Si, S, and Fe are treated.
The available model grids allow for a variable Fe abundance.

\begin{figure}[t]
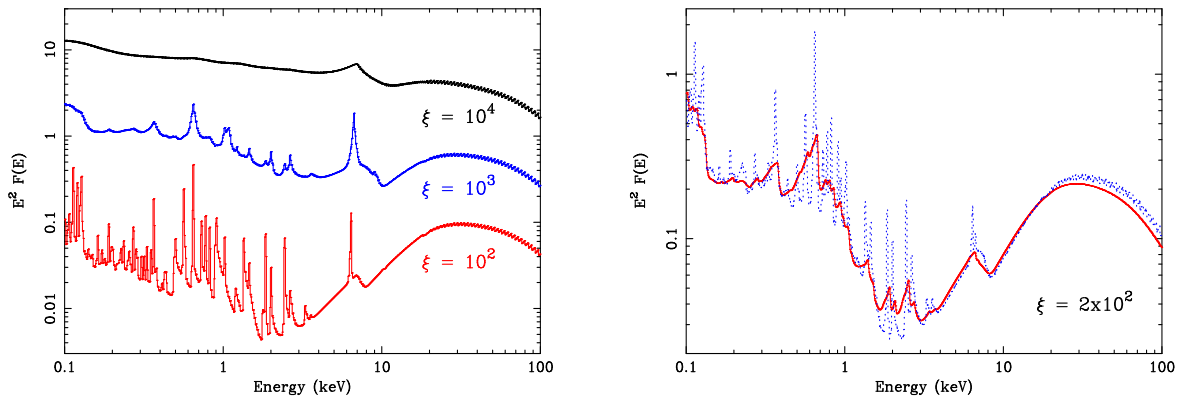
 
\begin{center} 
\hbox{
\psfig{figure=rossreflion.ps,width=0.45\textwidth,height=0.32\textwidth,angle=-90}
\hspace{0.8cm}
\psfig{figure=refblur.ps,width=0.45\textwidth,height=0.32\textwidth,angle=-90}
}
\vspace{-1cm} 
\end{center} 
\caption{\footnotesize { {\it Left panel}: Computed X--ray reflection
    spectra as a function of the ionization parameter $\xi$ (from the
    code by Ross \& Fabian 2005).  The illuminating continuum has a
    photon index of $\Gamma=2$ and the reflector is assumed to have
    cosmic (solar) abundances. {\it Right panel}: Relativistic effects
    on the X--ray reflection spectrum. We assume that the intrinsic
    rest--frame spectrum (dotted blue) is emitted in an accretion disc
    and suffers all the relativistic effects discussed above (see text
    for details). The relativistically--blurred reflection spectrum is
  shown in red.}}
\label{reflion} 
\end{figure}

In the left panel of Fig.~\ref{reflion} we show X--ray reflection
spectra produced by the code for three different values of the
ionization parameter (all other parameters being fixed). The
ionization parameter has clearly a large effect on the resulting
spectrum, most remarkably on the emission lines. For
$\xi=10^4$~erg~cm~s$^{-1}$ (top black) the surface layer is very
highly ionized and the only noticeable line is a highly
Compton--broadened Fe K$\alpha$ line peaking at 7~keV. The overall
spectral shape closely resembles that of the illuminating continuum (a
cut--off power law with photon index $\Gamma=2$). For
$\xi=10^3$~erg~cm~s$^{-1}$ (middle blue) the strong Fe K$\alpha$ line
is dominated by the Fe~{\footnotesize{XXV}} intercombination line,
while K$\alpha$ lines from the lighter elements emerge in the
0.3-3~keV band. Further reducing the ionization parameter to
$\xi=10^2$~erg~cm~s$^{-1}$ gives rise to a spectrum dominated by
emission features below 3~keV atop a deep absorption trough. The most
prominent feature is the Fe K$\alpha$ one at 6.4~keV. No residual
Compton broadening of the emission lines is visible.

In the right panel of Fig.~\ref{reflion} we show two versions of a
model with an ionization parameter of $\xi=2\times 10^2$~erg~cm~s$^{-1}$.
The blue one is the X--ray reflection spectrum in the absence of any
relativistic effect, whereas in red we show the
relativistically--blurred version of the same model, i.e.\ the spectrum
that is observed if reflection occurs from the accretion disc. All
sharp spectral features of the unblurred spectrum (blue) are broadened
by the relativistic effects explained above which makes it difficult
to identify clear emission lines in the soft spectrum.  Below about
2~keV the situation is often complicated by the presence of
absorption/emission features due to photoionized gas complicating the
soft part of the spectrum (see Fig.~\ref{wabs}, left panel). Thanks to
its strength, isolation, and to the fact that it occupies a region of
the X--ray spectrum relatively free from absorption, the Fe line is
however clearly seen. This is what makes this particular emission
feature a remarkable and unique tool that allows us to investigate the
dynamics of the innermost accretion flow via relativistic effects in
accreting BH systems.

\subsection{Dependence on the inner disc radius}

Einstein's equations imply the existence of an innermost radius within
which the circular orbit of a test particle in the equatorial plane is
no longer stable. This radius is known as the Innermost Stable
Circular Orbit (ISCO), sometimes referred to as the marginally stable
orbit (Bardeen, Press \& Teukolsky 1972). Beyond the ISCO, test
particles rapidly plunge into the BH on nearly geodesic orbits with
constant energy and angular momentum. By making the standard
assumption that the accretion disc is made of gas particles in
circular, or nearly circular, orbital motion, the disc extends down to
the ISCO, and emission from the plunging region is ignored. We shall
discuss further this assumption in the next Section.

\begin{figure}[t]
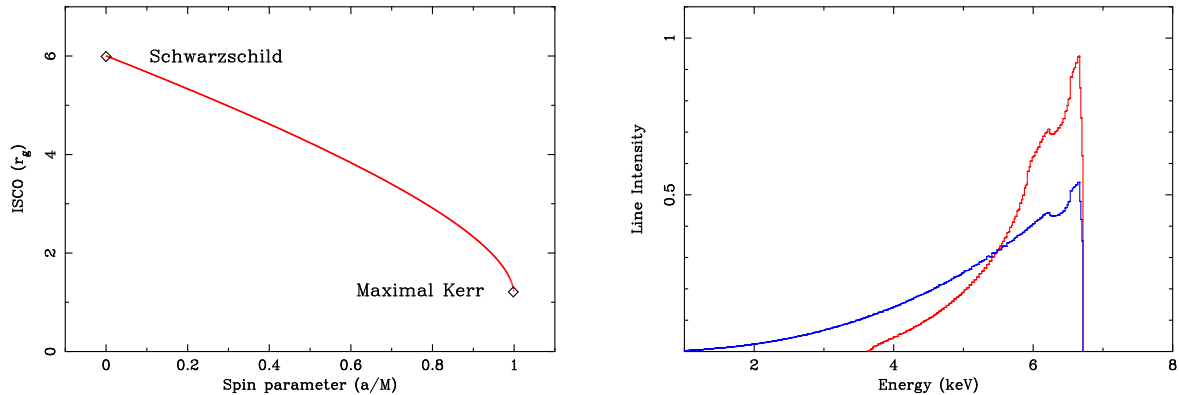
 
\begin{center} 
\hbox{
\psfig{figure=isco.ps,width=0.45\textwidth,height=0.32\textwidth,angle=-90}
\hspace{0.8cm}
\psfig{figure=diskline.ps,width=0.45\textwidth,height=0.32\textwidth,angle=-90}
}
\vspace{-1cm} 
\end{center} 
\caption{{\footnotesize{ {\it Left panel}: The dependence of the ISCO
      from the BH spin. We consider the allowed spin range from
      $a/M=0$ (Schwarzschild solution) to $a/M=0.998$ (Maximal Kerr
      solution). For these extremal cases, the ISCO is located at
      $6~r_g$ ($a/M=0$) and $\simeq 1.24~r_g$ ($a/M=0.998$). {\it
        Right panel}: The line profiles dependence from the inner disc
      radius is shown for the two extremal cases of a Schwarzschild BH
      (red, with inner disc radius at $6~r_g$) and of a Maximal Kerr
      BH (blue, with inner disc radius at $\simeq 1.24~r_g$) }}}
\label{risco}
\end{figure}

The actual radius of the ISCO depends on the BH spin parameter $a/M$
which can take any value for $0$ (Schwarzschild BH) to $1$ (maximally
spinning Kerr BH). As pointed out by Thorne (1974), the maximal value
for the spin parameter is likely to be about $0.998$ and we shall
refer to that case as ``Maximal Kerr''. The dependence of the ISCO
from the BH spin parameter is illustrated in the left panel of
Fig.~\ref{risco}.  The ISCO lies at $6~r_g$ from the centre for a
Schwarzschild BH, and at $\simeq 1.24~r_g$ for a Maximal Kerr one,
where $r_g = GM/c^2$ is the gravitational radius. It should be
stressed that the dependence is quite steep. If emission from say
$3~r_g$ can be detected the implied BH spin would be $a/M >
0.78$. Notice that accretion naturally causes a BH to spin up provided
the disc angular momentum is oriented as that of the hole (the
possible history of the spin of massive BHs is discussed e.g.\ by
Volonteri et al 2005 and references therein).

The inner boundary of the accretion disc, i.e.\ the location of the
ISCO, has a large impact on the shape of the line profile, especially
on its broad red wing. This is shown in the right panel of
Fig.~\ref{risco} where the cases of a Schwarzschild (red) and Maximal
Kerr BH (blue) are computed. The line is much broader in the Kerr case
because the smaller inner disc radius implies that the line photons
are suffering stronger relativistic effects (such as gravitational
redshift) which is visible in the resulting line profile. To summarise,
the detection and modelling of a relativistic broad Fe line via X--ray
observations of accreting BH potentially provides crucial information
on the system inclination, the radial efficiency of the coronal hard
X--ray emission, and also on one of the two parameters that
characterise the Kerr solution, i.e.\ the BH spin.

% Several AGN and black hole X-ray binaries show a clear very broad iron
% line which is strong evidence that the black holes are rapidly
% spinning. Detailed analysis of these objects shows that the emission
% line is not significantly affected by absorption and that the source 
% variability is principally due to variation in amplitude of a
% power-law. Underlying this is a much less variable,
% relativistically-smeared, reflection-dominated, component which
% carries the imprint of strong gravity at a few gravitational radii.
% The strong gravitational light bending in these regions then explains
% the power-law variability as due to changes in height of the primary
% X-ray source above the disc.  The reflection component, in particular
% its variability and the profile of the iron line, enables us to 
% study the innermost regions around an accreting, spinning, black
% hole. 

\subsection{The ISCO and its relation to black hole spin}
\label{sec:isco}

A key issue in using the profile of a broad iron line to determine BH
spin is whether the derived inner radius equals the ISCO, or not.
Reynolds \& Begelman (1997) pointed out that matter in the plunge
region within the ISCO can also contribute to reflection and so could
confuse spin determination.  The matter in the plunge region is
however moving inward rapidly and so has a low density. This means
that the ionization parameter is high, which changes the ability to
produce a detectable iron line. Young, Ross \& Fabian (1998) showed in
the key case of MCG--6-30-15 that the emission (and absorption)
produced by reflection in the plunge region does not resemble the
observed line shape. The issue of matter falling within the ISCO was
further considered by Gammie (1999) and Krolik (1999); see also Agol
\& Krolik (2000) with the inclusion of magnetic fields which can
connect the outer part of the plunge region with the disc and slow
matter down (also Krolik \& Hawley 2002). In fact, Dov\v{c}iak, Karas
\& Yaqoob (2004) have shown that the difference in the flow direction
has a negligible effect on the profile shape.

We suspect that the effect of magnetic fields on the iron-line
appearance is small. The ionization parameter
can be written as
\begin{equation}
\xi = 3\times 10^7 f (v_{\rm rad}/c) \quad \mathrm{erg~cm~s}^{-1} \ , 
\end{equation}
where $f$ is the volume filling factor of the flow which is falling
radially inward at velocity $v_{\rm rad}$ (Fabian \& Miniutti, in
preparation). Since $f$ and $v_{\rm rad}$ are unlikely to be much
smaller than 0.1 over most of the plunge region, it is most unlikely
that a strong iron line, requiring $\xi\sim
100$--$1000$~erg~cm~s$^{-1}$, can be produced. We suspect that the
inner radius determined from iron line studies cannot originate from
far within the ISCO and that the inner disc radius $r_{\rm in}$ is
never more than one $r_{\rm g}$ or so within the ISCO.

\section{The broad relativistic Fe line of MCG--6-30-15}

The X-ray spectrum of the bright Seyfert 1 galaxy \mcg\ ($z=0.00775$)
has a broad emission feature stretching from below 4~keV to about
7~keV. The shape of this feature, first clearly resolved with \asca\
by Tanaka \et (1995), is skewed and peaks at about 6.4~keV. This
profile is consistent with that predicted from iron fluorescence from
an accretion disc inclined at 30~deg extending down to within about 6
gravitational radii ($6r_{\rm g} = 6GM/c^2$) of a BH. During the
\asca\ observation the line appeared to be occasionally redshifted to 
lower energies, implying emission from even smaller radii and
therefore strongly suggesting that the BH in this object is rapidly
spinning (Iwasawa et al 1996b). The broad iron line of \mcg\ has been
detected by all major X--ray mission (\asca, \xte, \sax, \axaf, and
\xmm) with very similar results. 

\begin{figure}[] 
\begin{center} 
\hbox{
\psfig{figure=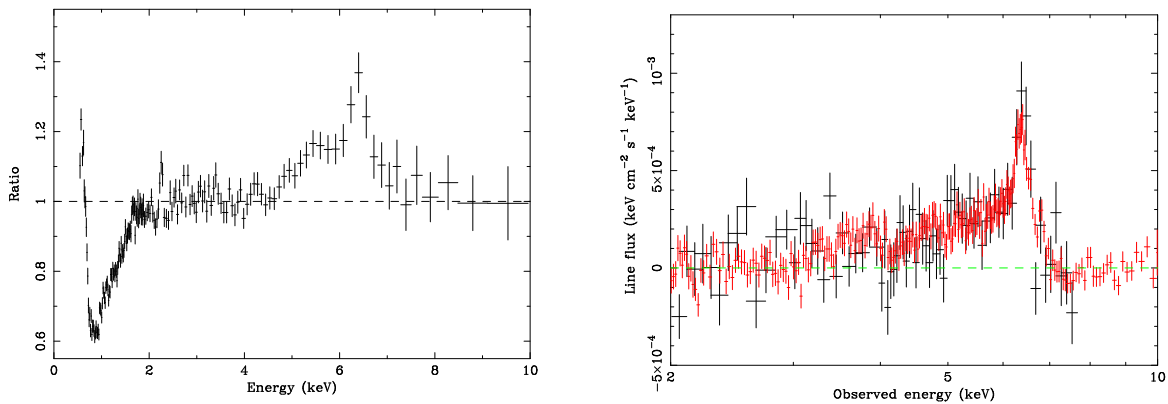,width=0.49\textwidth,height=0.36\textwidth,angle=-90}
\hspace{0.2cm}
\psfig{figure=mcg6youngline.ps,width=0.45\textwidth,height=0.32\textwidth,angle=-90}
\hspace{1.0cm}

} \vspace{-1cm} \end{center} \caption{\footnotesize {\it Left
  panel}: Ratio to a power law model of the 1994 ASCA S0 spectrum of
\mcg. The broad Fe line is clearly detected, together with absorption
from photoionized gas below 2~keV. {{\it Right panel}: Superposition
  of the broad Fe line profile of \mcg\ from the long \xmm\ (red) and
  \axaf\ (black) observations. The observations are non--simultaneous
  but the line profiles superimpose very well (Young et al 2005).}}
\label{asca}
\end{figure}

In Fig.~\ref{asca} we show the line profile obtained at different
times with three different detectors. In the right panel, we show the
broadband spectrum (as a ratio with a power law model) of the S0
detector on board of \asca\ during the 170~ks observation in 1994
(Iwasawa et al 1996b). In the right panel, we show the line profile
from non--simultaneous \xmm\ (red, 320~ks in 2001) and \axaf\ (black,
522~ks in 2004) observatories (Fabian et al 2002a; Young et al 2005).
The large effective area provided by \xmm\ confirmed earlier results
by Iwasawa et al (1996b; 1999) showing that the broad red wing of the
line extends down to about 3~keV implying a spin parameter $a/M >0.93$
(Dabrowski et al.\  1997; Reynolds et al.\  2004). Here we report in
some detail results from the \xmm\ observations of \mcg\ (Wilms et al
2001; Fabian et al 2002a; Fabian \& Vaughan 2002; Vaughan \& Fabian
2004; Reynolds et al 2004) mentioning also the most important
consequences of the recent long \axaf\ observation (Young et al 2005).

\begin{figure}[]
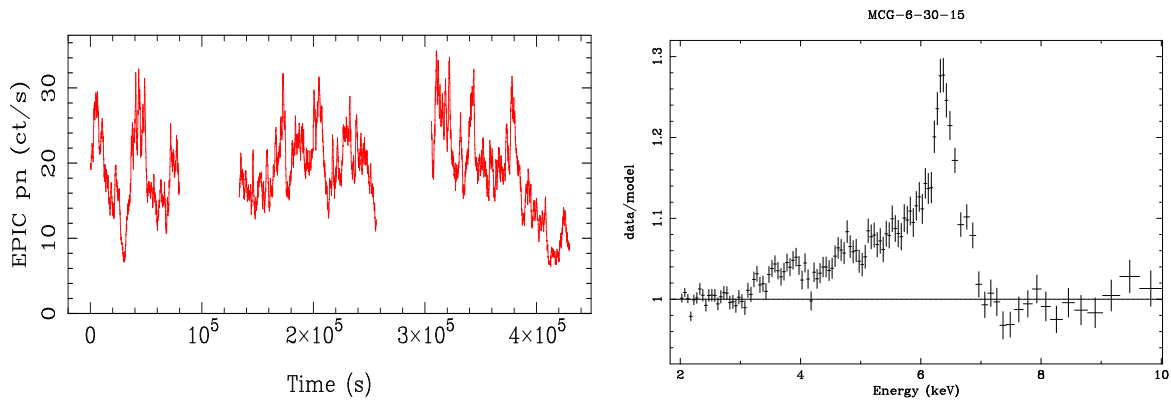
 
\begin{center} 
\hbox{
\psfig{figure=lca.ps,width=0.48\textwidth,height=0.30\textwidth,angle=-90}
\hspace{0.2cm}
\psfig{figure=mcg6_linebw.ps,width=0.45\textwidth,height=0.32\textwidth,angle=-90}
\hspace{1.0cm}

} \vspace{-1cm} \end{center} \caption{\footnotesize {\it Left
  panel}: The broadband light curve of the long \xmm\ observation in
2001. Three satellite's orbits have been devoted to study \mcg\ in
2001, for a total exposure of about 320~ks. {{\it Right panel}: The
  broad iron line in \mcg\ from the \xmm\ observation in 2001 (Fabian
  et al.\ 2002a) is shown as a ratio to the continuum model. }}
\label{mcgline}
\end{figure}

The 2001 \xmm\ observation was very long (about 320~ks corresponding
to three full satellite orbits) and simultaneous with a \sax\ one,
providing unprecedented spectral coverage with \xmm\ being more
sensitive than \sax\ in the 0.2--10~keV band, and \sax\ extending the
data set up to 200~keV. The broadband \xmm\ light curve for the three
orbits is shown in the left panel of Fig.~\ref{mcgline}. In the right
panel of the same figure we show the observed broad Fe line profile
from the time--averaged spectrum. We discuss here our best description
of the X--ray spectrum of \mcg, focusing on the broad Fe line, and
later review alternatives to our modelling which, however, all prove
largely unsatisfactory when tested against the data.

The broadband spectrum of \mcg\ is best described by a
relativistically blurred reflection spectrum modified by absorption by
photoionized gas below about 2~keV (Fabian et al 2002a). For the
X--ray reflection model, we used the code by Ross \& Fabian (1993;
2005) allowing the photon index $\Gamma$, the ionization parameter
$\xi$, and the relative strength of the reflection spectrum with
respect to the power law continuum free to vary. To account for
Doppler and gravitational effects, the reflection spectrum is
convolved with a relativistic kernel computed using a modified version
of the code by Ari Laor (1991) which is appropriate for the Maximal
Kerr case.  However, the inner disc radius is not fixed to the ISCO
but can be larger.  Therefore, if the inner disc radius can be
inferred from the data, information on the spin can in principle be
obtained by making use of the ISCO--spin relation (see
Fig.~\ref{risco}). The emissivity profile has the general form of a
broken power law with index $q_{\rm{in}}$ from the inner disc radius
to a break radius $r_{\rm{br}}$ and $q_{\rm{out}}$ outwards.

The best--fitting parameters indicate a weakly ionized disc with
ionization parameter $\xi < 30$~erg~cm~s$^{-1}$ and a strong
reflection fraction\footnote{The reflection fraction 
($R$) is defined as
  the relative normalization of the reflection component with respect
  to the illuminating continuum. $R=1$ means that the primary source
  subtends a solid angle $\Omega/2\pi = 1$ at the disc. Values larger
  than $1$ imply that the disc is seeing more illuminating continuum
  that any observer at infinity and therefore that the primary source
  is anisotropic and preferentially shines towards the disc.} of about
$R\simeq 2.2$, a result which is fully confirmed by the \sax\ data
where the Compton hump around 20--30~keV is clearly seen. Fe is
measured to be three times more abundant than standard (solar). The
parameters of the relativistic blurring are the disc inclination
$i=33^\circ \pm 1^\circ$, the inner disc radius $r_{\rm{in}}=1.8 \pm
0.1~r_g$, and the emissivity parameters $q_{\rm{in}} = 6.9\pm 0.6$,
$q_{\rm{out}}=3.0\pm 0.1$, and $r_{\rm{br}}=3.4\pm 0.2$ (Vaughan \&
Fabian 2004). The measured value of the inner disc radius implies
emission from far beyond the ISCO of a Schwarzschild BH ($6~r_g$) and
therefore suggest the BH in \mcg\ is rapidly spinning.  If we take the
measured value of $r_{\rm{in}}$ as a measure of the ISCO, we can
constrain the BH spin to be $a/M=0.96 \pm 0.01$, providing one of the
most remarkable indications so far for the astrophysical relevance of
the Kerr solution to the Einstein's equations. The inferred emissivity
profile indicates a standard emissivity from about $3.4~r_g$ to the
outer disc boundary. However, a steep profile is required in the
innermost few gravitational radii, indicating that most of the
accretion power is released in the region where General Relativity is
in the strong field regime. Tapping of black holes spin by magnetic
fields in the disc is a strong possibility to account for the peaking
of the power so close to the hole (Wilms et al.\ 2001; Reynolds et al.\
2004; Garofalo \& Reynolds 2005).

\subsection{Alternatives to a relativistic line}

The claim that iron line studies are probing the strong field regime
of General Relativity just few gravitational radii from the BH is a
bold one, and should always be tested again alternative models.  In
this spirit, we discuss here alternatives to the broad relativistic Fe
line in the X--ray spectrum of \mcg\ (see also Fabian et al 1995;
Fabian et al 2000). 

The broad Fe line profile seen in the right panel of
Fig.~\ref{mcgline} comprises two main components: a relatively narrow
core peaking around 6.4~keV, and an extended broad red wing from about
3~keV to 6~keV. If the broad red wing is not due to a relativistic
line but instead to some spectral curvature independent of any
relativistic effect, we are left with the line core only. If this is
emitted from far the BH (such as from the torus), it has to be narrow
and unresolved at the \xmm\ resolution (about 100~eV). However, the
line core itself is unambiguously resolved by \xmm. When modelled by a
Gaussian emission line, the line core width is $\sigma =
352^{+106}_{-50}$~eV, much larger than the spectral resolution
(Vaughan \& Fabian 2004).
Considering a blend of lines from different Fe ionization states as
responsible for the broadening produces an unacceptable fit.
Therefore, even ignoring for a moment the broad red wing of the line,
the relatively narrow line core itself already suggest that the line
is emitted in a fast orbiting medium close to the central BH and
excludes any strong contribution from a reflector located at the
parsec (and even sub--parsec) scale such as the torus and/or the Broad
Line Region. A small contribution is possible and likely, but at the
level of 15 per cent maximum of the line core.

\begin{figure}[]
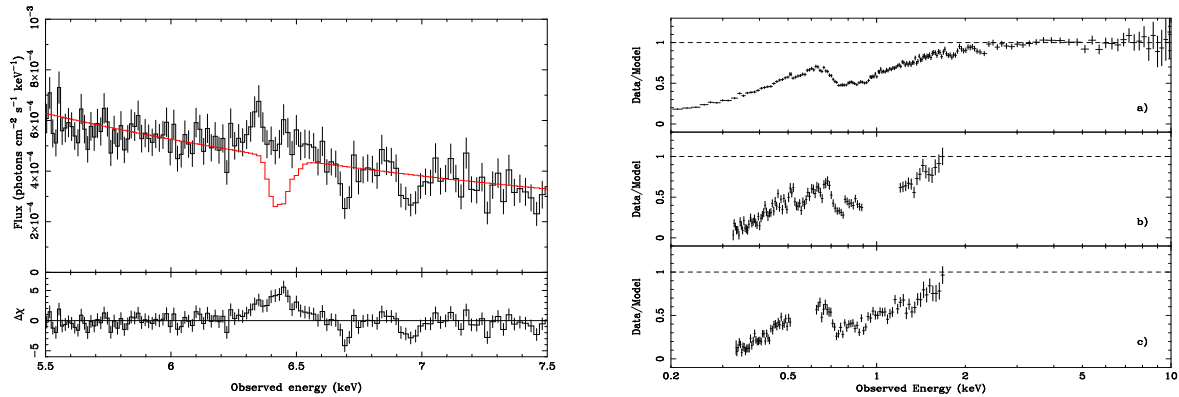
 
\begin{center} 
  \hbox{
    \psfig{figure=young7.ps,width=0.45\textwidth,height=0.32\textwidth,angle=-90}
    \hspace{0.8cm}
    \psfig{figure=combined_absorption_2.eps,width=0.45\textwidth,height=0.31\textwidth,angle=-90}
  } \vspace{-1cm} \end{center} \caption{\footnotesize {{\it Left
      panel}: The \axaf\ HEG spectrum (black) is overlaid to the
    best--fit ionized absorption model of the broad red wing of the
    iron line. The ionized absorber produces a curved spectral shape
    that approximates the broad red wing, but predicts iron absorption
    features in the 6.4--6.6~keV range (rest--frame) which are
    inconsistent with the data (Young et al 2005). {\it Right panel}:
    The difference spectrum obtained by subtracting the low flux state
    spectrum from the high flux state one, plotted as a ratio to a power law in
    the 3--10~keV band. The top panels refers to the EPIC--pn data,
    the two lower panels to the RGS gratings on \xmm\ (Turner
    et al 2003).}}
\label{chandra}
\end{figure}

As for the broad red wing of the line below 6~keV (see right panel of
Fig.~\ref{mcgline}) several attempts have been made to explain the
spectral shape without invoking relativistic effects. The broad red
wing of the line has  been modelled by different emission
components, i.e.\ a thermal component, a blend of broadened emission
lines (other than Fe), and complex absorption. In the latter model,
which is probably the only serious one, the continuum passes through a
large column of moderately ionized gas which can cause significant
curvature above the Fe~L edge (about 0.7~keV) up to about 5--6~keV,
therefore with the potential of reproducing the broad red wing of the
line. The gas must be sufficiently highly ionized to avoid undetected
excessive opacity in the soft band, but not so highly ionized to lose
of all its L--shell electrons. The best--fit ionized absorption model
to both \xmm\ and \axaf\ data can reproduce the red wing of the line
(though the fit statistic is much worse than when the data are fitted
with a relativistic line). However, the model predicts a clear complex
of narrow absorption lines between 6.4~keV and 6.7~keV. This is
inconsistent with the high resolution \axaf\ data as shown in the left
panel of Fig.~\ref{chandra} from Young et al (2005). We conclude that
ionized absorption cannot reproduce the broad red wing of the
relativistic Fe line.

A neutral partial covering model (which would be consistent with
\axaf\ data) is also ruled out because it is inconsistent with the
high energy data provided by \xte\ and \sax, in particular the
unambiguous requirement for a strong reflection component (Reynolds et
al 2004; Vaughan \& Fabian 2004). As a further test, a partial
covering model was added to the best--fit model with a relativistic
line to see if the derived parameters (most importantly the inner disc
radius providing information on the BH spin) are robust. No
differences were noticed and $r_{\rm{in}}$ remains at $1.8\pm 0.1~r_g$
(Vaughan \& Fabian 2004; Young et al 2005). An ionized partial
covering model can not be simply ruled out, but \axaf\ data constrain
the covering fraction to be less than 5 per cent, insufficient to
produce enough spectral curvature in the relevant energy band,
therefore requiring the presence of a broad relativistic line as for
the neutral partial covering model above. As discussed in the next
Section, the presence of complex absorption above 3~keV is also
strongly (and in our opinion, unambiguously) ruled out by spectral
variability analysis.

\subsection{A puzzling spectral variability}

The X-ray continuum emission of \mcg\ is highly variable (see Vaughan,
Fabian \& Nandra 2003; Vaughan \& Fabian 2004; Reynolds et al.\ 2004;
McHardy et al 2005 for some recent analyses) as is also clear from the
light curve in Fig.~\ref{mcgline}. Several different methods are used
to explore the spectral X--ray variability in accreting BH sources.
One such a method is provided by the analysis of the so--called
difference spectrum, i.e.\ the spectrum that can be obtained by
subtracting a low flux state spectrum from a high flux one (Fabian \&
Vaughan 2003). In this way any component that remained constant within
the two flux levels is effectively removed from the spectrum which
therefore carries information on the variable components plus
absorption only. The \xmm\ 2001 observation was split into 2 different
flux states according to the light curve (left panel of
Fig.~\ref{mcgline}) and two spectra were extracted as representative
of the high and low flux states. The low flux spectrum was then
subtracted from the high flux one.

In the right panel of Fig.~\ref{chandra} we show the resulting
difference spectrum as a ratio to a power law fitted in the 3--10~keV
band (from Turner et al 2003). The EPIC--pn data are shown in the
upper panel, while the high resolution \xmm--RGS gratings are plotted in
the middle and lower panels.  It is clear from the figure, that the
variable component is very well described by the power law in the
3--10~keV band and is affected by absorption from photoionized gas
only below 3~keV. This demonstrates in a model--independent way that
the X--ray variability in \mcg\ is due to a relatively steep
($\Gamma\simeq 2.2$) power law component, and that the amplitude
changes of the continuum in \mcg\ are due to this power law.

The difference spectrum also clarifies in a model--independent way
that there is no subtle additional absorption above 3~keV that might
influence the broad red wing of the relativistic Fe line.  Absorption
by either neutral or ionized gas has to show up in the difference
spectrum (as seen below 3~keV) but is not detected in the Fe band
above 3~keV. This is in our opinion conclusive on the reality of the
broad red wing of the Fe line. The case for the relativistic Fe line
in \mcg\ therefore looks even more robust now than before.

It should be stressed that the difference spectrum, unlike the true
spectrum of the source, does not exhibit any emission feature (neither
narrow nor broad) in the Fe band. This implies that the relativistic
Fe line in \mcg\ is much less variable than the power law continuum
and has therefore been removed by the subtraction. As a test, the \xmm\
2001 exposure was divided into 10~ks long chunks and spectral fits
performed. Direct spectral analysis on the 10~ks spectra confirms that
the variability in \mcg\ closely follows a two--component model
comprising a Power Law Component (hereafter PLC) with constant slope
of about $\Gamma=2.2$ and highly variable normalization (a factor
3--4) and a Reflection--Dominated Component (RDC) that carries the
broad Fe line and whose variability is confined within 25 per cent
(Fabian \& Vaughan 2003; Vaughan \& Fabian 2004) confirming previous
results (Shih, Iwasawa \& Fabian 2002; Matsumoto et al 2003).

\begin{figure}[]
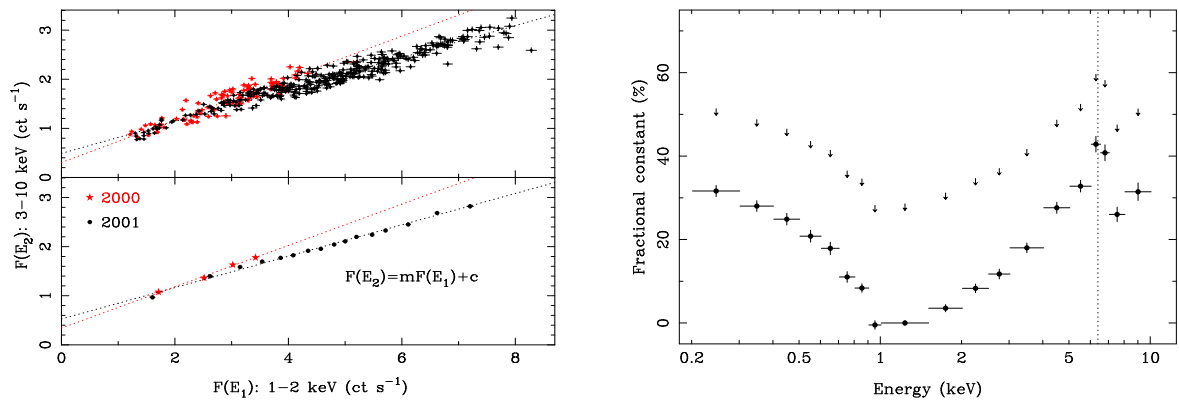
 
\begin{center} 
  \hbox{
    \psfig{figure=mcgff.ps,width=0.45\textwidth,height=0.32\textwidth,angle=-90}
    \hspace{0.8cm}
    \psfig{figure=mcgfrac.ps,width=0.45\textwidth,height=0.32\textwidth,angle=-90}
  } \vspace{-1cm} \end{center} \caption{\footnotesize {{\it Left
      panel}: Flux--flux plot for the 2000 (red) and 2001 (black)
    \xmm\ observations. The 3--10~keV count rates are plotted versus
    the simultaneous 1--2~keV count rates. The upper panel shows the
    raw data, the lower is a binned version of the same plot. {\it
      Right panel}: The fractional spectrum of the constant component
    of \mcg\ constructed from the y--axis offset in flux--flux plots
    generated in different bands. The spectrum strongly resembles that
    of a reflection component, with a prominent contribution at
    6.4~keV (vertical dotted line). Figures from Vaughan \& Fabian (2004).}}
\label{mcgff} 
\end{figure}

A further powerful model--independent tool to explore the spectral
variability is provided by flux--flux plot analysis which consists of
plotting the count rates in two different energy bands against each
other (Taylor, Uttley \& McHardy 2003). The largest variability in
\mcg\ occurs in the 1--2~keV band which is therefore chosen as the
reference energy band.  In the left panel of Fig.~\ref{mcgff} we show
the count rates in the 3--10~keV band plotted against the simultaneous
count rates in the 1--2~keV band (from Vaughan \& Fabian 2004). Black
points refer to the 2001 long \xmm\ observation, red ones to the
previous shorter \xmm\ observation in 2000 (Wilms et al 2001). The
Figure is divided into two panels, the lower one being simply a binned
version. It is clear that a very significant correlation is
found in both observations. The scatter in the unbinned version is a
manifestation of the different shape of the power spectral density in
the two bands and lack of coherence. The scatter is removed in the
binned version which is most useful for fitting purposes.

The relationship is clearly linear which is an unambiguous sign that
the variability in both bands is dominated by a spectral component
which has the same spectral shape and varies in normalization only.
However, extrapolating the linear relation to low count rates leaves a
clear offset on the y--axis. This strongly suggests the presence of a
second component that varies very little and contributes more to the
3--10~keV than to the 1--2~keV band. In other words, the flux in each
band is the sum of a variable and a constant component, with the
constant component contributing more to the harder than softer
energies. From the previous discussion on the difference spectrum, the
variable component is easily identified with a power law with constant
slope $\Gamma=2.2$ dominating the variability by variation in
normalization only. Moreover, we already know that the reflection
component varies little because no Fe line is seen in the difference
spectrum and because of the result of direct time--resolved spectral
analysis. It is therefore likely that the constant component is
associated with the X--ray reflection spectrum, including the
relativistic Fe line.

Flux--flux plots can be produced for any chosen band by plotting the
relative count rates as a function of those in the reference band.
Each plot will produce a different y--axis offset which represents the
fractional contribution of the constant component to the spectrum in
that particular energy band. Therefore, by recording the value of the
offset in each band, the ``fractional spectrum'' of the constant
component can be obtained and the hypothesis that this is similar to a
X--ray reflection spectrum can be tested. The fractional spectrum of
the constant component is shown in the right panel of Fig.~\ref{mcgff}
(from Vaughan \& Fabian 2004; see Taylor, Uttley \& McHardy for an
similar result based on \xte\ data). The black circles are the result
obtained by assuming that the contribution of the constant component
in the 1--1.5~keV band is null, the arrows indicate the upper limits
(notice that the shape, which is the relevant here, does not change).
The fractional spectrum is soft below $\sim$~1~keV and hard at higher
energies with a strong contribution at 6.4~keV. This strongly resembles
a typical X--ray reflection spectrum, supporting our previous
hypothesis that the constant component in \mcg\ is indeed
reflection--dominated.

\subsection{Interpretation: the light bending model}

\begin{figure}[]
\begin{center}
\hbox{\hspace{0cm}
\psfig{figure=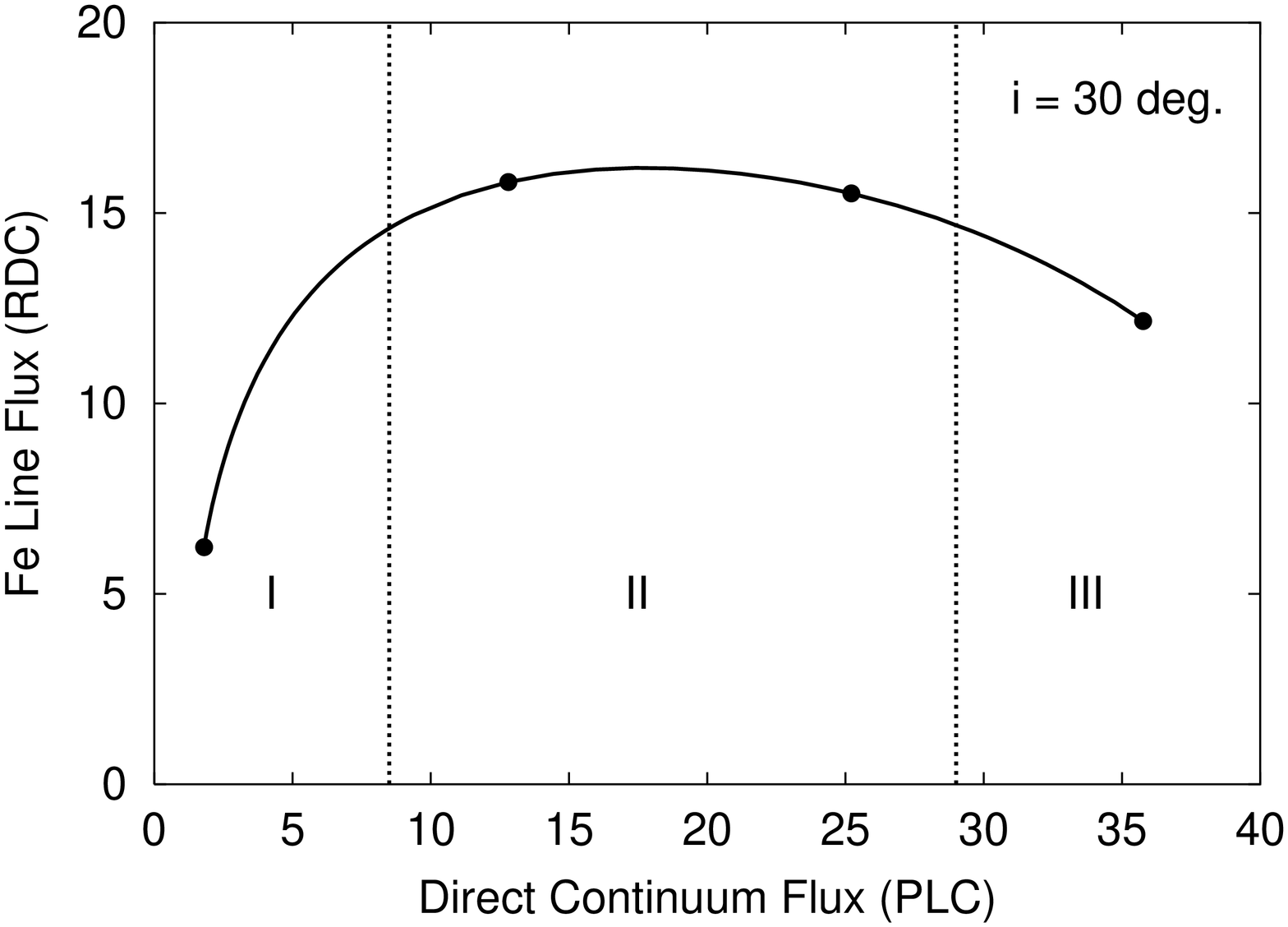,width=0.48\textwidth,height=0.34\textwidth,angle=-90} 
\hspace{0.4cm}
\psfig{figure=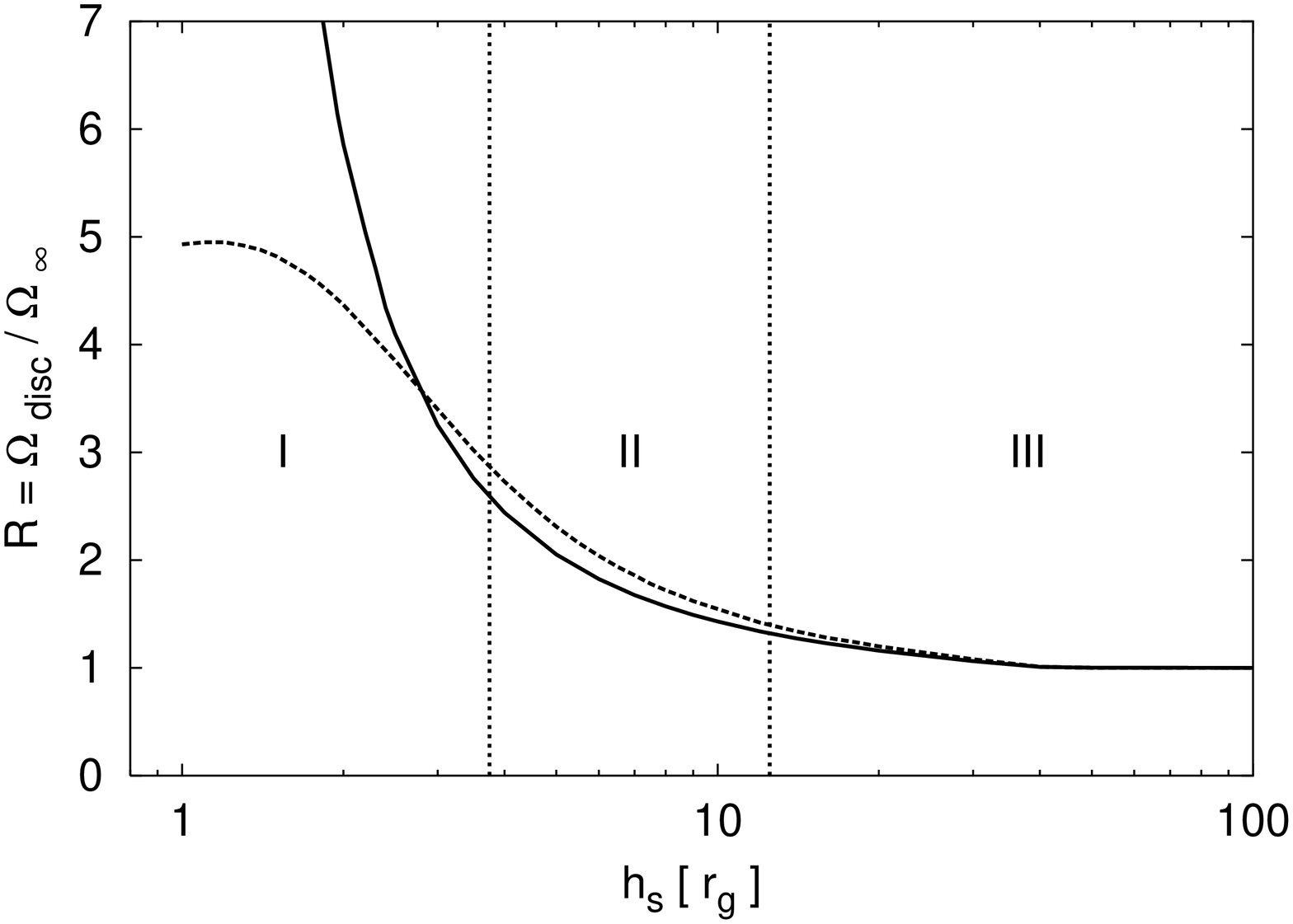,width=0.48\textwidth,height=0.34\textwidth,angle=-90}
} 
\vspace{-1cm} 
\end{center} 
\caption{\footnotesize {{\it Left panel}: The Fe line flux
    (representative of the whole RDC) as a function of the PLC flux
    observed at infinity. The variability is only due to changes in
    the primary source height. Black dots represent heights of
    $h=1,\,5,\,10,\,20~r_g$. Three regimes can be identified. In
    regime II the RDC variation is within 10 per cent while the PLC
    varies by about a factor 4. {\it Right panel}: The reflection
    fraction as a function of the source height (nearly the same if
    plotted versus the PLC flux). We show the cases of a source on the
    disc axis (solid) and one at $2~r_g$ from the axis.  } }
\label{lbending} 
\end{figure}

If the observed power law component (PLC) continuum drives the iron
fluorescence then the line flux should respond to variations in the
incident continuum on timescales comparable to the light-crossing, or
hydrodynamical time of the inner accretion disc (Fabian et al 1989;
Stella 1990; Matt \& Perola 1992; Reynolds et al.\ 1999). This
timescale ($\sim 100M_6$~s for reflection from within $10r_{\rm g}$
around a black hole of mass $10^6 M_6$~\Msun) is short enough that a
single, long observation spans many light-crossing times. This has
motivated observational efforts to find variations in the line flux
(e.g.\  Iwasawa \et 1996b, 1999; Reynolds 2000; Vaughan \& Edelson
2001; Shih, Iwasawa \& Fabian 2002; Matsumoto et al 2003). However,
the variability analysis discussed above implies that the
reflection--dominated component (RDC) does not follows in general the
PLC variations.

Explaining the relatively small variability of the RDC, compared with
that of the PLC, provides a significant challenge. The RDC and PLC
appear partially disconnected. Since however both show the effects of
the warm absorber they must originate in a similar location. As the
extensive red wing of the iron line in the RDC indicates emission
peaking at only a few gravitational radii ($GM/c^2$) we must assume
that this is indeed where that component originates. On the other
hand, this is also the region where most of the accretion power is
dissipated and therefore the PLC must originate in the corona above
the innermost disc as well, so that light--travel--time effects cannot
be invoked to explain the lack of response of the RDC to the PLC
variation.

We have developed a model for the variability of the PLC and RDC
components in accreting BH systems which is based on the idea that
since both components originate in the immediate vicinity of the BH,
they both suffer relativistic effects due to strong gravity (Fabian \&
Vaughan 2003; Miniutti et al 2003; Miniutti \& Fabian 2004). Some of
the effects of the strong gravitational field on the line shape and
intensity had already been explored by Martocchia, Matt \& Karas
(2000) and Dabrowski \& Lasenby (2001). We assume a geometry in which
a Maximal Kerr BH is surrounded by the accretion disc extending down
to the ISCO. The PLC is emitted from a primary source above the
accretion disc. Its location is defined by its radial distance from
the BH axis and its height (h) above the accretion disc.  Several
geometries have been studied and the results we present are
qualitatively maintained for any source location and geometry within
the innermost 4--5 gravitational radii from the black hole axis. The
main parameter in our model is the height of the primary source above
the accretion disc and the main requirement is for the source to be
compact. The primary PLC source could be physically realised by flares
related to magnetic reconnection in the inner corona, emission from
the base of a jet close to the BH, internal shocks in aborted jets,
dissipation of the rotational BH energy via magnetic processes, etc.
Any mechanism producing a compact PLC--emitting region above the
innermost region of the accretion flow would be relevant for our model
(e.g.\ Blandford \& Znajek 1977; Markoff, Falcke \& Fender 2001; Li
2003; Ghisellini, Haardt \& Matt 2004 and many others).

A fraction of the radiation emitted by the primary source directly
reaches the observer at infinity and constitutes the direct continuum
which is observed as the PLC of the spectrum.
The remaining radiation illuminates the accretion disc (or is lost
into the hole event horizon). The radiation that illuminates the disc
is reprocessed into the RDC. For simplicity, we assume that the
intrinsic luminosity of the primary source is constant. The basic idea
is that the relevant parameter for the variability of both the PLC and
the illuminating continuum on the disc (which drives the RDC
variability) is the height of the primary source above the accretion
disc and that the variability is driven by general relativistic
effects the most important of which is gravitational light bending.

As an example, if the source height is small (of the order of few
gravitational radii) a large fraction of the emitted photons is bent
towards the disc by the strong gravitational field of the BH enhancing
the illuminating continuum and strongly reducing the PLC at infinity,
so that the spectrum is reflection--dominated. If the source height
increases, gravitational light bending is reduced so that the observed
PLC increases.  Finally, if the height is very large so that light
bending has little effect, the standard picture of reflection models
with approximately half of the emitted photons being intercepted by
the disc and the remaining half reaching the observer as the PLC is
recovered.
\begin{figure}[]
\begin{center}
\hbox{\hspace{0cm}
\psfig{figure=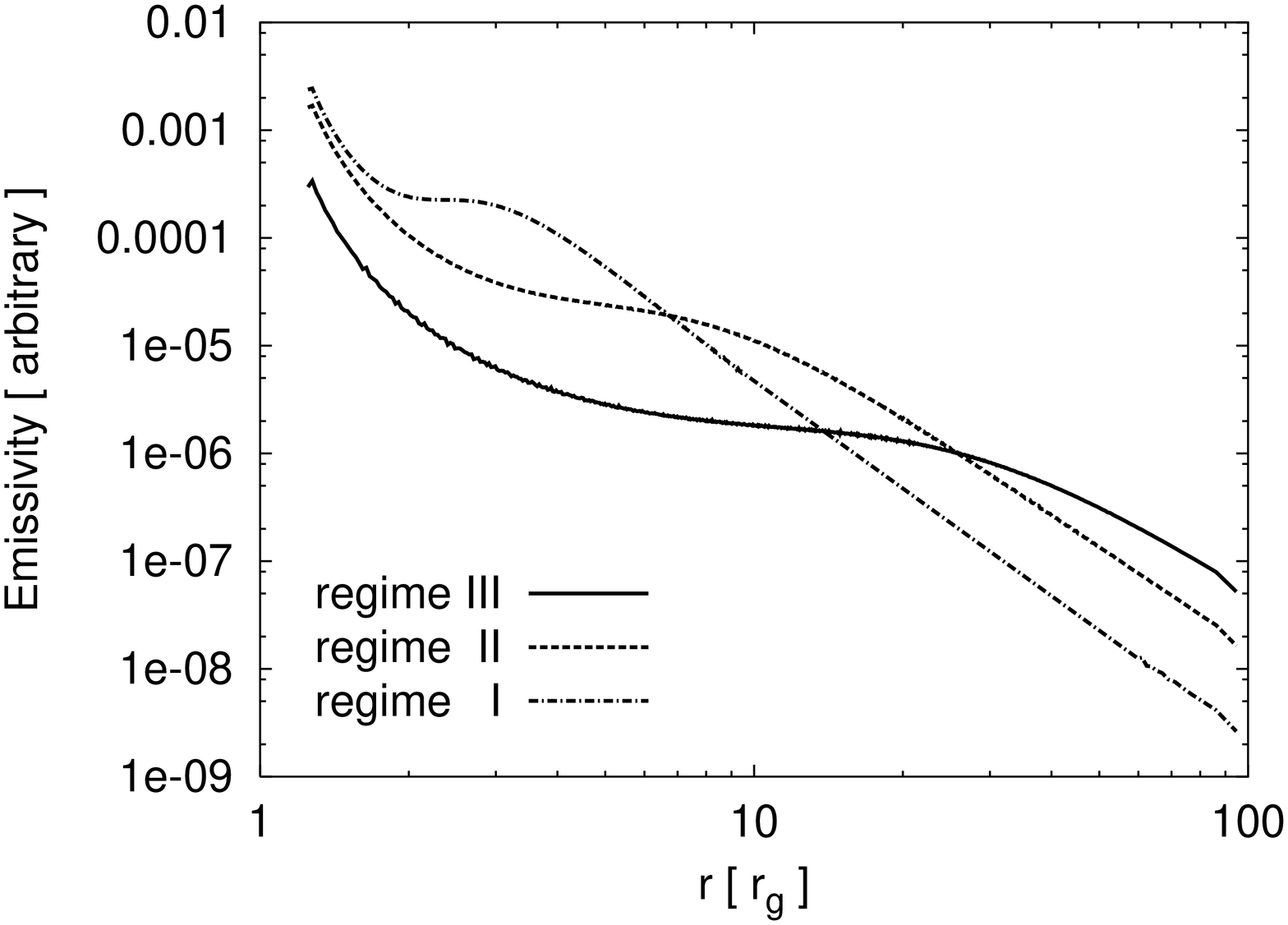,width=0.48\textwidth,height=0.34\textwidth,angle=-90} 
\hspace{0.4cm}
\psfig{figure=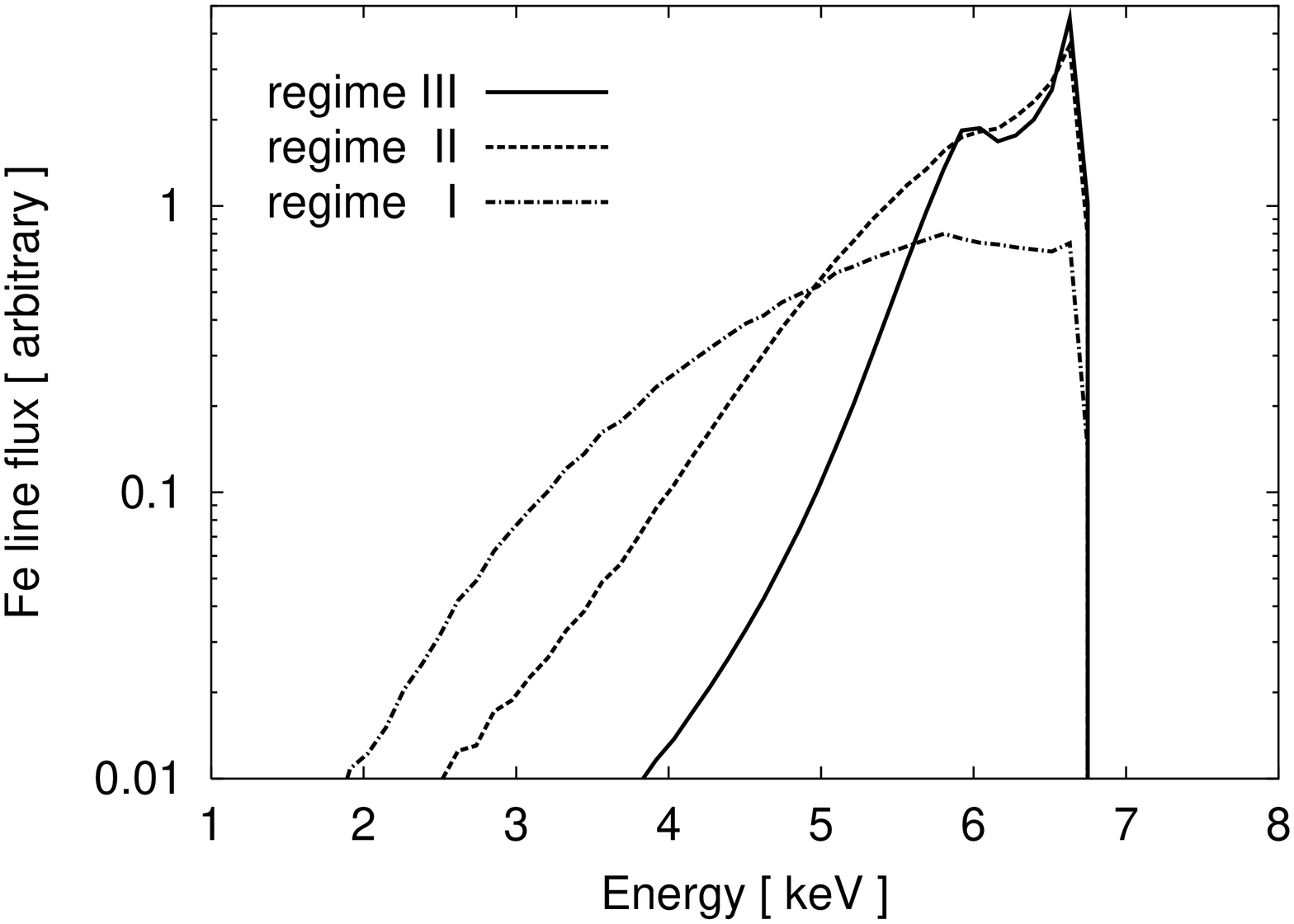,width=0.48\textwidth,height=0.34\textwidth,angle=-90}
} 
\vspace{-1cm} 
\end{center} 
\caption{\footnotesize 
{{\it Left panel}: Typical emissivity profiles computed in the light
  bending model for the three relevant regimes. The emissivity is
  steeper in the inner part of the disc and flattens outwards. 
{\it Right panel}: Typical line profiles in the three regimes. A log
scale is used to enhance the line profile differences. 
} }
\label{lbending2} 
\end{figure}

With this setup, we compute the PLC and Fe line flux (representative
of the whole RDC) as a function of the primary source height above the
disc (see Miniutti et al 2003; Miniutti \& Fabian 2004 for more
details). Our results are presented in the left panel of
Fig.~\ref{lbending} for the case of an accretion disc seen at an
inclination of $30^\circ$ (the relevant case for \mcg). The black dots
indicate different values for the source height ($h=1, 5, 10, 20~r_g$
from left to right). What the figure shows is that, once general
relativistic effects are properly accounted for, the broad Fe line
(and RDC) is not expected to respond simply to the observed PLC
variation anymore.  Three regimes can be identified (I, II, and III in
the figure). A correlation is seen only in regime I, where the primary
source height is between about $1$ and $3$--$4~r_g$. On the other
hand, if the primary source has a height between about $4~r_g$ and
$12~r_g$ (regime II) the broad Fe line (and RDC) variability is
confined within 10 per cent only, while the PLC can vary by a factor
$\sim$~4. We recall here that the observed variability of the RDC in
the 2001 \xmm\ observation is confined within 25 per cent despite
variations by a factor 3--4 in the PLC.

In the right panel of Fig.~\ref{lbending} we show the value of the
reflection fraction predicted by the model as a function of the source
height. When the source height is large, the standard value $R=1$ is
recovered. However, as the height decreases, light bending comes into
play increasing the number of photons bent towards the disc while 
simultaneously reducing the number of those able to escape from the
gravitational attraction. Thus, the reflection fraction increases
dramatically. Two cases are shown in the figure, one for a source on
the disc axis (solid), one for a ring--like configuration at $2~r_g$
from the black hole axis. Regime II, which seems most relevant so far,
is characterised by $R\simeq 2$. Quite remarkably this is precisely
the measured value in \mcg\ during the simultaneous \xmm--\sax\ 2001
observation we are discussing here.

Thus, if our model is qualitatively correct, it implies that \mcg\ was
observed in regime II by \xmm\ during the 2001 observation with most
of the accretion power being dissipated within $5~r_g$ in radius and
in the range $4$--$12~r_g$ in height above the disc. This is itself a
remarkable result implying that we are looking down to the region 
of spacetime where General Relativity exhibits itself at its best.
Since the model computes self--consistently the disc illumination, the
emissivity profile can be computed. In regime II, it turns out to be
best described by a broken power law, very steep in the innermost
region of the disc and flattening outwards, as observed in \mcg\ (see
left panel of Fig.~\ref{lbending2}). We computed fully relativistic
line profiles with our model and compared them with the data
(Fig.~\ref{lbending2}, right panel). The line profiles obtained from
regime II provide a very good match to the \xmm\ data
(Miniutti et al 2003).

\begin{figure}[]
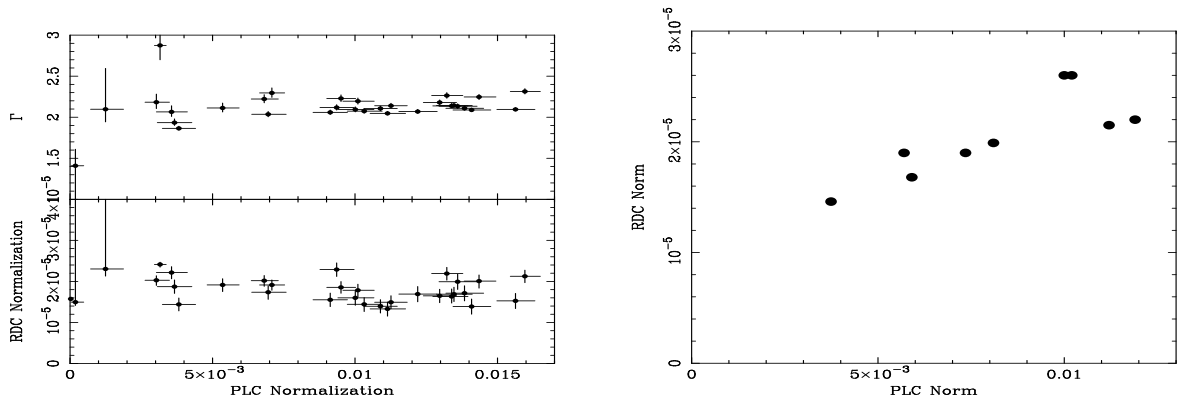

\begin{center}
\hbox{\hspace{0cm}
\psfig{figure=normplot.ps,width=0.45\textwidth,height=0.30\textwidth,angle=-90} 
\hspace{0.8cm}
\psfig{figure=rdcplc.ps,width=0.45\textwidth,height=0.32\textwidth,angle=-90}
} 
\vspace{-1cm} 
\end{center} 
\caption{\footnotesize 
{{\it Left panel}: PLC photon index (top) and RDC normalization as a
  function of the PLC normalization for the 2001 observation.
{\it Right panel}: RDc normalization as a function of the PLC
normalization for the 2000 \xmm\ observation.
} }
\label{00and01} 
\end{figure}

As is clear from Fig.~\ref{lbending} (left panel), the model
predicts a correlation between RDC and PLC at low flux levels.
Fortunately, the \xmm\ 2000 observation (that we did not discuss so
far) caught \mcg\ in a slightly lower flux state than the 2001 one.
The flux state is not remarkably lower, but some indications of the
general behaviour can still be inferred.  The details of the spectral
analysis of the 2000 observation can be found in Wilms et al (2001) and
Reynolds et al (2004). Here we only focus on the variability of the two
components. If the source did not change dramatically its properties
between the two observations, we should observe a somewhat correlated
variability between the RDC and the PLC.  A comparison between the two
observations in this sense is shown in Fig.~\ref{00and01} where
results of spectral analysis performed on 10~ks chunks for both
observations are presented. In the right panel, the PLC slope $\Gamma$
and RDC normalization are shown as a function of the PLC normalization
for the higher flux 2001 \xmm\ observation. The relevant panel is the
bottom one showing the constancy of the RDC despite the large PLC
variation. In the right panel of the same figure, we show the RDC vs.
PLC normalization for the lower flux 2000 \xmm\ observation. As
already reported by Reynolds et al (2004), a clear correlation is seen
between the two components in this lower flux observation. This result
was not known yet when the light bending model was being developed and
provides unexpected support for the overall picture, matching one of
the main predictions of the model.

The puzzling variability of the RDC and PLC, the large value of the
reflection fraction, and even the emissivity and line profiles all
seem to indicate the relevance of the light bending model for \mcg.
Given the very strong case for Fe line emission from the relativistic
region in \mcg, it would be a rather surprising coincidence if strong
relativistic effects were not responsible for some, if not all, of the
behaviour discussed above. We conclude by stressing that in all cases
in which broad Fe lines are detected strong relativistic effects on
the X--ray emission and variability cannot be ignored and have to be
included in theoretical models. The light bending model is probably
crude and incomplete, but represents a first significant step in this
direction.

\subsection{Occasional Fe line variability}

\begin{figure}[t!]
  \begin{center} \hbox{ \hspace{-2.5cm}
      \psfig{figure=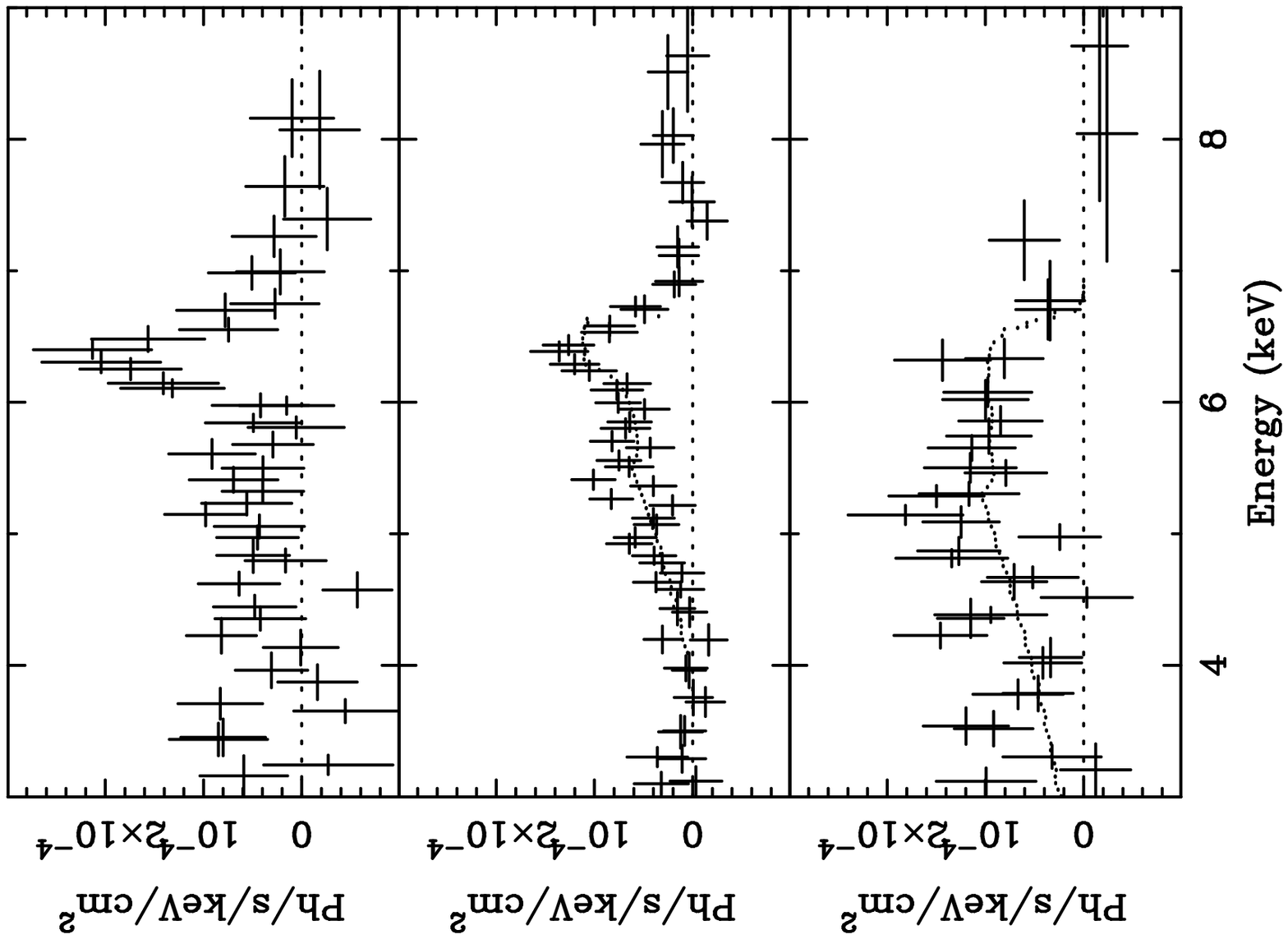,width=0.85\textwidth,height=0.45\textwidth,angle=-90}
      \hspace{-3.5cm}
      \psfig{figure=gioki.ps,width=0.45\textwidth,height=0.4\textwidth,angle=-90}
    } \vspace{-1cm} \end{center} \caption{\footnotesize {{\it Left
        panel}: The observed variations of the line profile during the
      1994 \asca\ observation (Iwasawa et al 1996). From top to bottom the high, medium,
      and low flux states profiles are shown.  {\it Right panel}: The
      changes in the line profile predicted by the light bending model
      for the observed flux states.  }} \label{ki96}
\end{figure}

As mentioned, the RDC and broad Fe line in \mcg\ is almost constant
despite large variation in the PLC. However, it exhibits occasionally
marked variations in the line profile such as during the 1994 \asca\
observation (Iwasawa et al 1996b). Spectra were extracted in a high,
medium, and very low flux state, the difference in count rate between
the low and high flux being about a factor 4. The resulting line
profiles are shown in the left panel of Fig.~\ref{ki96} where, from
top to bottom, the line profile is for the high, medium and low flux
state respectively. The general trend appears to be that the line is
more peaked in high than low flux states and becomes much broader as
the flux decreases. In particular, in the lowest flux state, no peak
is visible at 6.4~keV and the line is very broad and redshifted.
During the so--called deep minimum (bottom--left panel in Fig.\ref{ki96})
the best--fitting parameters for the line profile indicate that
emission is coming from within $6~r_g$ providing the first evidence
for the presence of a Kerr BH in \mcg, later confirmed by higher
quality \xmm\ data.

In the framework of the
light bending model, low flux states correspond to situations in which
the primary source is very close to the BH thereby illuminating very
efficiently the inner disc. In higher flux states, the source has a
larger height, and the outer disc starts to be illuminated as well.
Therefore, the model predicts very broad and redshifted lines in low
flux states (see also Fig.~\ref{lbending2}, right panel) while a peak
at 6.4~keV due to emission from the outer regions of the disc appears
at higher flux levels. In the right panel of Fig.~\ref{ki96} we show
three profiles obtained in the light bending model for three different
heights of the primary source ($h=6,\,4,\,2~r_g$ from top to bottom).
The profiles match the observed data rather well. It should be stressed that
the PLC variation between $h=2~r_g$ (low flux) and $h=6~r_g$ (high
flux) is predicted to be a factor 5, very similar to the factor 4 of
the data. 

\begin{figure}
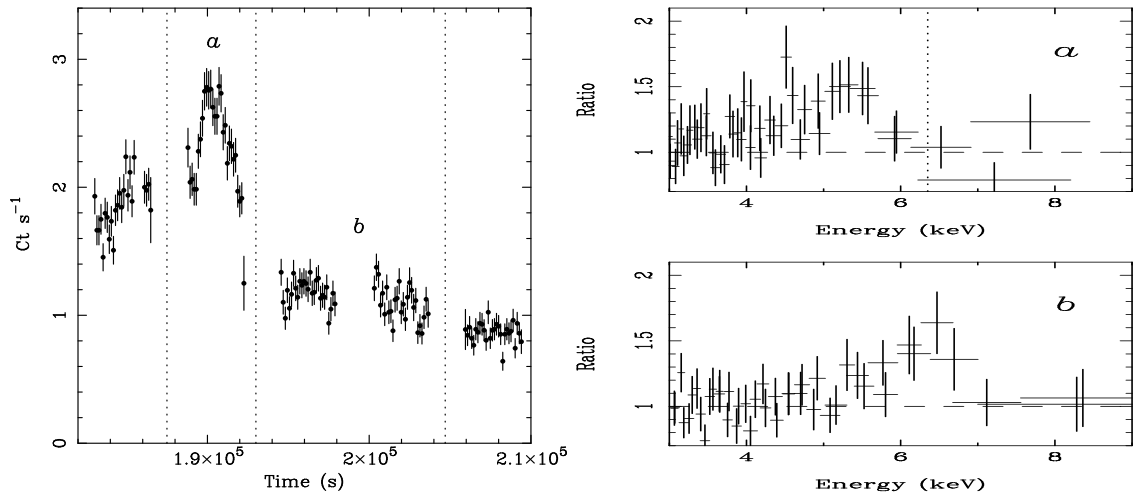

\begin{center} 
  \hbox{
    \psfig{figure=ki99lc.ps,width=0.46\textwidth,height=0.4\textwidth,angle=-90}
%    \hspace{-0.2cm}
    \psfig{figure=ki99lines.ps,width=0.46\textwidth,height=0.4\textwidth}
  } \vspace{-1cm} \end{center} \caption{\footnotesize {{\it Left
      panel}: The \asca\ light curve around the bright flare (Iwasawa
et al 1999). The two
    time--intervals relevant for the right panel of the figure are
    here defined.  {\it Right panel}: During the flare (a) the
    broad line is redshifted well below the Fe K$\alpha$ rest--frame
    energy (6.4~keV, vertical line). When the flare ceased (b) the
    broad line recovered to the ordinary profile.  }} \label{ki99}
\end{figure}

Further remarkable evidence for  variability of the Fe line on
short--timescales is provided by the 1997 \asca\ observation (Iwasawa
et al 1999). During a bright flare, the Fe line profile changed
dramatically. This is shown in Fig.~\ref{ki99} where the light curve
and line profiles in two relevant time--intervals are shown. One
possibility to explain such dramatic change is that a bright
corotating flare appeared at about $5~r_g$ from the BH. Given the low
BH mass in \mcg\ (about $10^6~M_\odot$) the flare has enough time to
complete more than one orbit during the time--interval `a', therefore
resulting effectively in a ring--like emitting structure which
provides a good fit to the data (Iwasawa et al 1999).

\section{Other AGN: Seyfert1 and NLS1 galaxies}

The case for a broad relativistic Fe line is very strong in \mcg. In
this source, the best--fitting parameters imply that Fe has about
three times the solar abundance, the disc is lowly ionized, and the
RDC is particularly strong with respect to the PLC ($R\simeq 2.2$).
These conditions make it easier to detect the relativistic line in
this AGN than in others.  Indeed, it should be stressed that if these
conditions were not met in \mcg, the detection of a relativistic Fe
line would be much more difficult, challenging present X--ray
observatories. A discussion on the detectability of broad Fe line in
less favourable conditions is deferred to Section~\ref{sec:broadfree}
below.

\begin{figure}[]
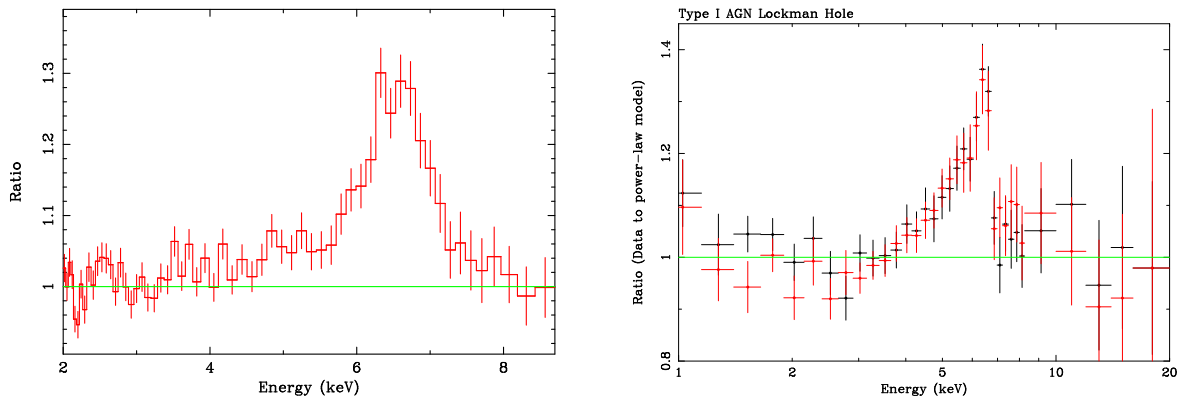
 
\begin{center} \hbox{
\psfig{figure=iras18325.ps,width=0.45\textwidth,height=0.32\textwidth,angle=-90}
\hspace{0.8cm}
\psfig{figure=alina_new.ps,width=0.45\textwidth,height=0.32\textwidth,angle=-90}
}
\vspace{-1cm} 
\end{center} 
\caption{\footnotesize {{\it Left panel}: Ratio of the spectrum of
    IRAS~18325--5926 to a power law for \xmm\ data. {\it Right panel}:
    Ratio plot of the mean unfolded spectrum for type--1 AGN in the
    Lockman Hole with respect to a power law (Streblyanska et al 2005). \xmm\ EPIC--pn and MOS
    detectors are used.  }}
\label{iras} 
\end{figure}

Here we just show another interesting case of a broad Fe line,
presented in Fig.~\ref{iras}. In the left panel, we plot the ratio of
the \xmm\ data of IRAS~18325--5926 with respect to a power law model
fitted in the 2--10~keV band. The Fe line in IRAS~18325--5926 is
clearly broad and exhibits a strong red wing extending down to about
4~keV (Iwasawa et al 1996a, Iwasawa et al 2004; Iwasawa et al, in
preparation). Some other AGN in which resolved Fe lines have been
reported in the past are discussed in Nandra et al (1997a,b; 1999);
Bianchi et al (2001); Lamer et al (2003); Longinotti et al 2003;
Balestra et al (2004); Porquet et al (2004b); Jim\'enez--Bail\'on et
al (2005), the list being non--exhaustive. Some of the Fe lines in the
list are not extreme and do not imply that the reflector extends down
below the ISCO of a non--rotating BH, thereby not constraining the BH
spin. The line width in these objects is however inconsistent with
emission from a very distant reflector such as the torus and/or the
Broad Line Region and seems often inconsistent with being a blend of
numerous narrower lines.

It should be stressed that not all broad lines are clear and
unambiguous as in \mcg, and other interpretations of the data are
often possible. A possible scenario is that complex absorption by
large columns of highly ionized matter produces a curvature in the
spectrum that resembles that of a broad Fe line (e.g.\ Turner et al
2005). As mentioned, this was conclusively excluded for \mcg\ but is a
possible alternative to relativistic lines in other cases. The
recently launched Astro--E2 mission has enough effective
area and energy resolution in the relevant Fe band to test this
scenario by detecting or not the absorption features associated with
it. High energy data will also be crucial because they can indicate
the presence/lack of strong reflection components through the
detection of the Compton hump therefore suggesting which
interpretation is preferable. 

The 770~ks \xmm\ observation of the Lockman Hole field provides also
some interesting information. The Lockman Hole is a special field in
the sky because of extremely low Galactic absorption. It is therefore
an ideal field to study faint AGN as a population and their
contribution to the X--ray background. The field comprises 53 type--1
and 41 type--2 AGN with known redshifts. Streblyanska et al (2005)
have obtained the mean spectrum of type--1 AGN by stacking together
the individual spectra and correcting for the source redshift (see
also Brusa et al 2005 for a spectral stacking results from the Chandra
Deep Fields). The ratio of the data to a power law model for the mean
type--1 AGN spectrum is shown in the right panel of Fig.~\ref{iras}
and exhibits a broad emission feature peaking at 6.4~keV. A prominent
broad red wing is also visible. The observed equivalent width is
larger than the average value for bright nearby Seyfert 1 galaxies but
similar for example to \mcg\ and might indicate that Fe is
overabundant with respect to solar values. Shemmer et al (2004) have
suggested that the metallicity in AGNs is correlated with the mass
accretion rate (close to the Eddington limit), which is in turn
related to the AGN luminosity. It is possible that the large
equivalent widths measured here are due to high metallicity in distant
and luminous objects. The average line profile is best--fitted with an
inner disc radius of $3~r_g$. This may be an indication that most of
the AGNs contain a Kerr BH at their very centre which would have
serious implications for the dominant BH growth mechanisms and, more
broadly, on cosmology and galaxy evolution.

\subsection{1H~0707--495: a key to the nature of NLS1?}

Narrow Line Seyfert 1 galaxies tend to show steep soft X-ray spectra,
extreme X--ray variability, and sometimes broad iron emission
features. One extreme such object is 1H\,0707-495 which has a marked
drop in its spectrum above 7~keV. This is either an absorption edge
showing partial-covering in the source (Boller et al.\ 2002; Gallo et
al 2004; Tanaka et al 2004) or the blue wing of a massive, very broad,
iron line (Fabian et al.\ 2002b, 2004). Here we explore in some detail
this latter possibility as an example of what could be a more general
picture relevant to many other objects. 1H\,0707--495 has been
observed twice by \xmm\ with a deep spectral drop at 7~keV 
discovered in the first observation. The drop had shifted to 7.5~keV
in the second observation two years later.

\begin{figure}[]
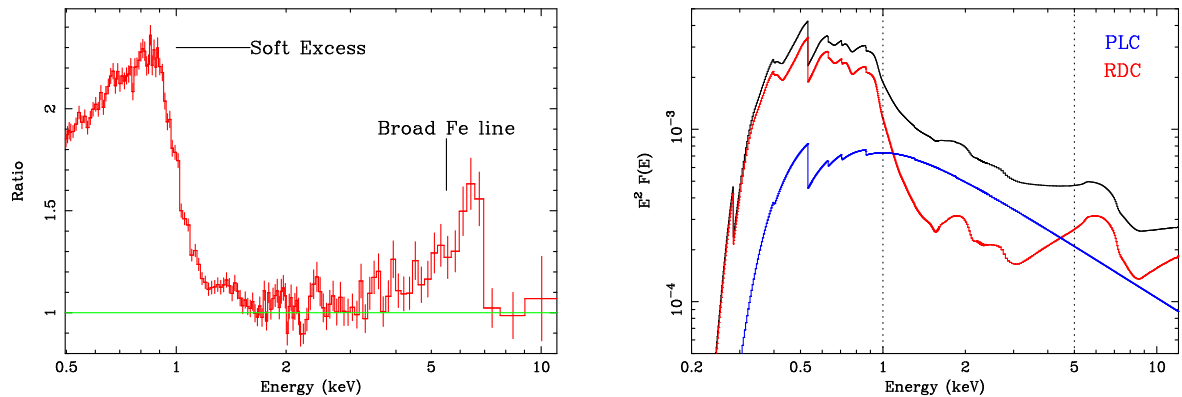
 
\begin{center} \hbox{
\psfig{figure=1h07line.ps,width=0.45\textwidth,height=0.32\textwidth,angle=-90}
\hspace{0.8cm}
\psfig{figure=1h07model.ps,width=0.45\textwidth,height=0.32\textwidth,angle=-90}
}
\vspace{-1cm} 
\end{center} 
\caption{\footnotesize {{\it Left panel}: Ratio of the spectrum of the
NLS1
1H0707 to a power-law fitted between 2 and 4~keV and above 7.5~keV.
{\it Right panel}: Spectral decomposition of 1\,H0707--495 in terms of a
variable power-law (blue) and a blurred reflection component
(red). Vertical lines separate the regions in which one of the two
components dominates the spectrum.
}}
\label{1h07line} 
\end{figure}

In the left panel of Fig.~\ref{1h07line} we show the ratio of the data
(second observation) to a simple power law model fitted between 2 and
4~keV and above 7.5~keV. A large skewed emission feature is seen, very
similar to that of \mcg\ (see the right panel of Fig.~\ref{mcgline})
together with a steep soft excess below 1--2~keV. We have fitted the
broadband 0.5--10~keV spectrum with a simple two--component model
comprising a PLC and a relativistically blurred RDC obtaining an
excellent description of the entire data set. The data require Fe to
have super--solar abundance (3 times the solar value) and the
reflector to be ionized at the level of $\xi\simeq
650$~erg~cm~s$^{-1}$. The X--ray reflection spectrum comes from the
innermost regions of the disc with inner disc radius measured at
$2.3~r_g$ and steep emissivity profile (Fabian et al 2004). In
comparison, the emissivity was even steeper during the first
observation which explains why the marked spectral drop (that we
interpret as the combination of the reflection edge and blue peak of
the Fe line) was at lower energy (7~keV in the first and 7.5~keV in the
second observation). The shift is due to gravitational redshift which
is more effective in lowering the energy of sharp features when the
emissivity is steeper (i.e.\ the first observation).

The best--fit model is
shown in the right panel of Fig.~\ref{1h07line}. The solution implies
that the spectrum is almost completely reflection--dominated, i.e.\ the
PLC illuminates much more efficiently the disc than the observer at
infinity. This seems therefore to be a promising source in which to
test further the light bending model that successfully reproduces the
spectral shape and variability of the key source \mcg. In fact the
model predicts the existence of reflection--dominated states in regime
I and part of regime II. 

\begin{figure}[] 
\begin{center} \hbox{
\psfig{figure=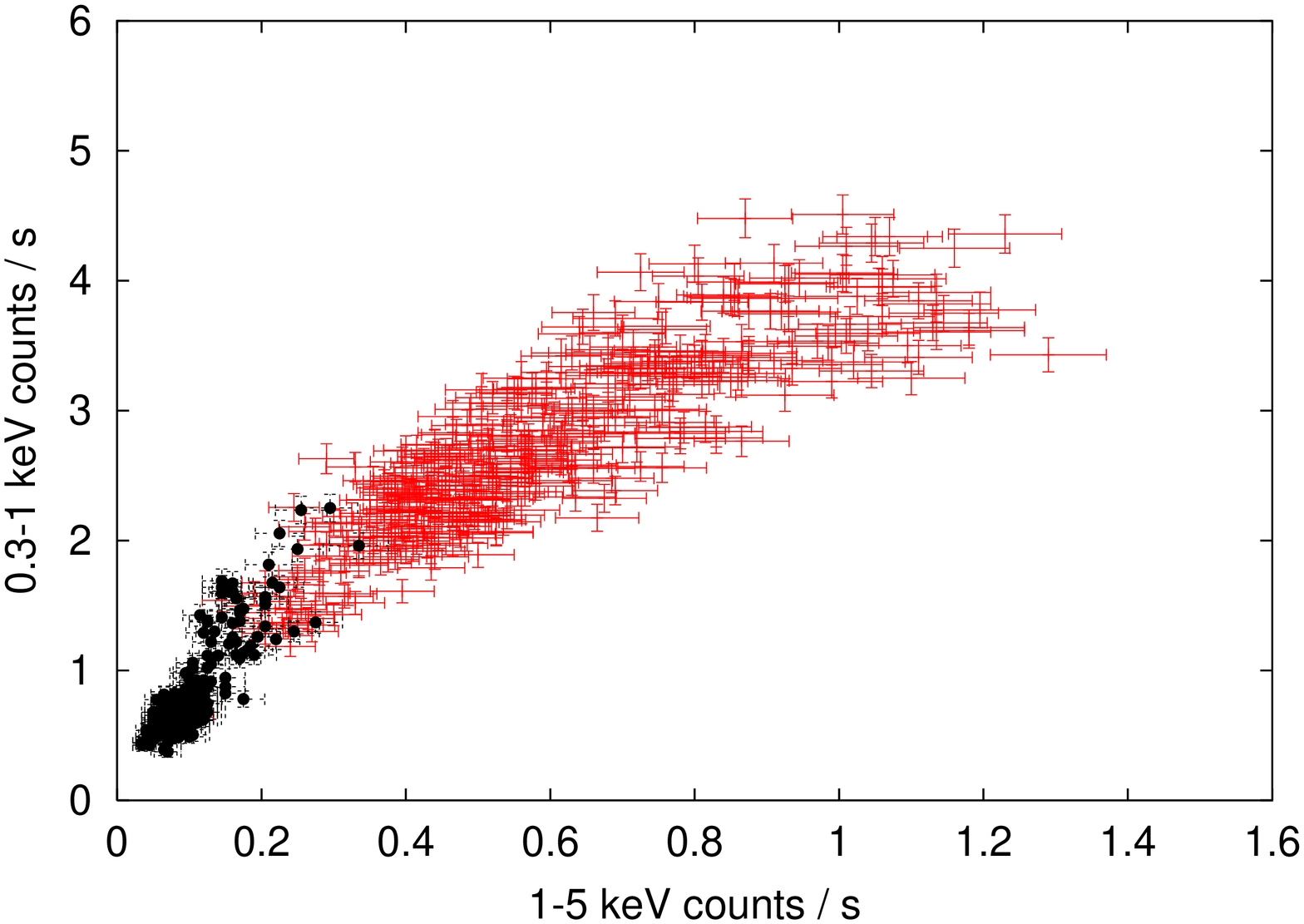,width=0.45\textwidth,height=0.34\textwidth,angle=-90}
\hspace{0.8cm}
\psfig{figure=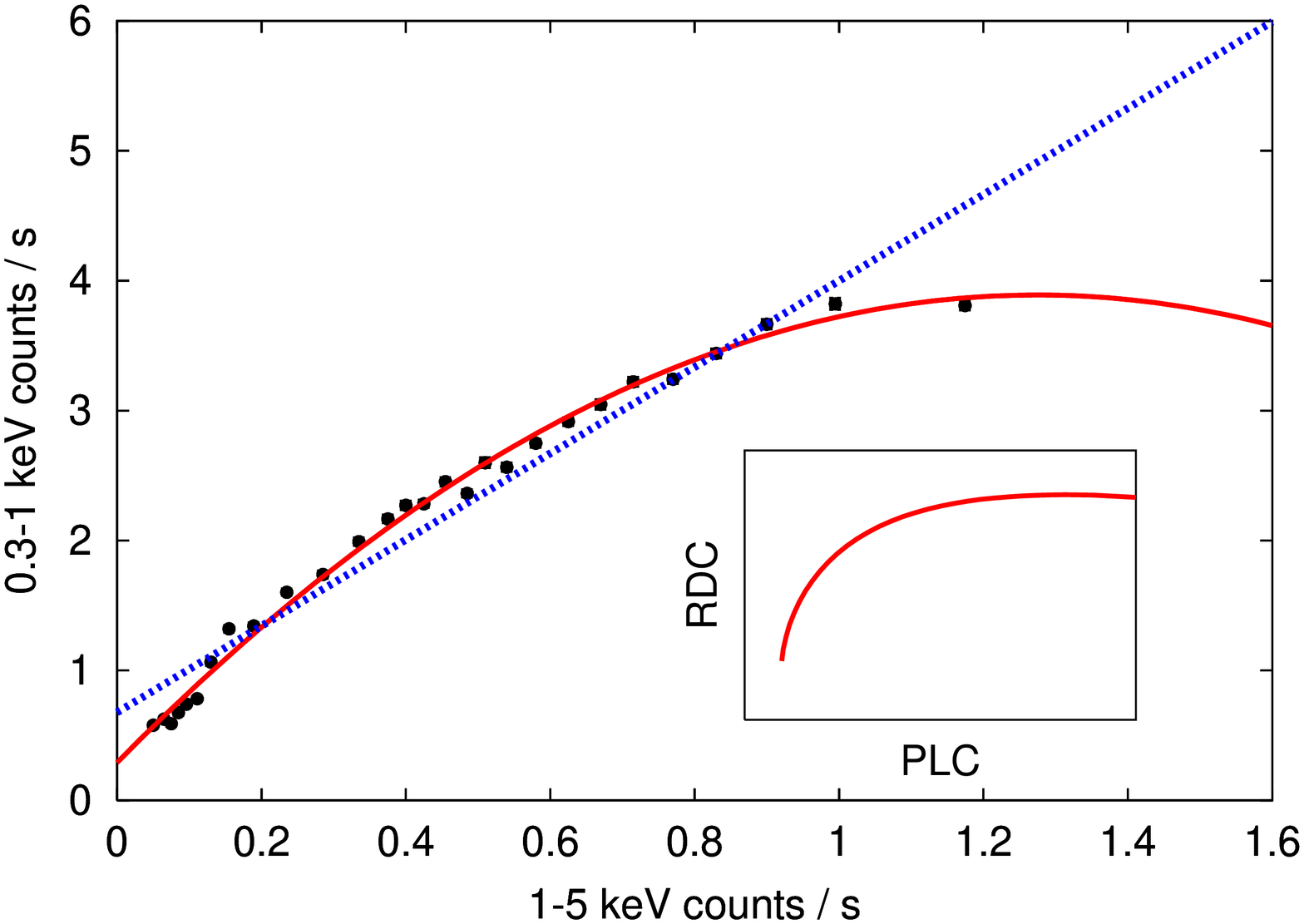,width=0.45\textwidth,height=0.34\textwidth,angle=-90}
}
\vspace{-1cm} 
\end{center} 
\caption{\footnotesize {{\it Left panel}: Flux--flux plot for the two
    observations of 1H\,0707--495. The 0.3--1~keV band is
    representative of the RDC, whereas the 1--5~keV band of the PLC
    (see right panel of Fig.~\ref{1h07line}).
{\it Right panel}: Binned version of the flux--flux plot. A linear
relation is a much worse fit than a curved one. In the insert panel,
we show the prediction of the light bending model.
}}
\label{1h07ff} 
\end{figure}

From the right panel of Fig.~\ref{1h07line}, the RDC dominates the
spectrum below about 1~keV and above 5~keV, whereas the PLC dominates
in the intermediate 1--5~keV band (the vertical lines in the figure
separate the different energy bands).  As a zeroth--order
approximation, we can then consider the 1--5~keV flux as
representative of the PLC, while the 0.3--1~keV flux is associated
with the RDC. If so, by plotting the 0.3--1~keV count rate as a
function of the 1--5~keV one, we could infer the RDC correlation with
the PLC. This is shown in the left panel of Fig.~\ref{1h07ff} for both
the first (black) and second (red) observations. There is clearly a
correlation that, however, is not linear. This is more
clearly shown in the binned version of the same plot, showed in the
right panel of Fig.~\ref{1h07ff}. The correlation is 
non--linear, and a curved relationship is preferred by the data. In
the insert, we show the prediction of the light bending model which
seems to catch the overall behaviour surprisingly well. The insert is
appropriate for an accretion disc seen at $30^\circ$, while the disc
inclination in 1H\,0707-495 is likely more $50^\circ$. As shown in
Miniutti \& Fabian (2004) the light bending model predicted behaviour
for larger inclination is even more similar to that seen in the data.

We conclude by noting that, quite remarkably, a relativistically
blurred reflection-dominated model describes well the XMM-Newton
spectrum of 1H\,0707-49 over the entire observed energy band (for both
observations). Strong gravitational light bending around a Kerr BH
seems to be the simplest explanation for the peculiar spectrum of this
source and its remarkable variability. More detailed analysis of the
data sets is given in Fabian et al (2002b; 2004) where further support
in favour of the light bending model is given. The same scenario
described above is also relevant for 1H~0419-577 (Fabian et al 2005)
and NGC~4051 (Ponti et al 2005). An alternative description in terms
of partial covering is provided by Boller et al (2002); Gallo et al
(2004); Tanaka et al (2004). See also Pounds et al (2004a) for a
similar partial--covering interpretation of the low flux state of
1H~0419--577.

\section{Galactic Black Holes Candidates}

Black hole binaries are accreting systems containing a dark compact
primary with a mass larger than about $3~M_\odot$ and a
non--degenerate companion. These are generally refereed to as
confirmed BH binaries because the high mass of the primary excludes it
is a neutron star. Only 18 such systems are known in our Galaxy, but
they are thought to be representative of a large population of systems
(possibly around 30 millions of sources). The first to be discovered
was Cyg--X1, containing a BH of $6.9$--$13.2~M_\odot$ with a O/B star
as companion. In addition, many other X--ray sources in the Galaxy do
exhibit all the observed characteristics of BH binaries. Many of these
sources can not be claimed as secure BH binaries because the mass of
the primary is not well constrained, and they are generally referred
to as Galactic BH Candidates (GBHCs). 

BH binaries and GBHCs exhibit a fascinating phenomenology which
manifest itself in complex X--ray spectral and temporal properties. In
many respects, these sources can be thought as scaled--down versions
of AGN both powered by accretion into the central BH. It is not our
purpose here to explore the complexity of BH binaries and GBHCs, and
we refer the interested reader to more specialistic reviews such as,
for example, the one by McClintock \& Remillard (2004). Our interest
here is to focus on the broad relativistic Fe lines that have been
detected in many objects, sometimes suggesting that the central BH is
rapidly spinning. One such a case, the GBHC XTE~J1650--500, will be
presented in some detail, mainly because its 2001 outburst was covered
by three different X--ray detectors (\xte, \sax, \xmm ) and is
therefore particularly well characterized (though this is clearly not
the only source which benefits from such an extensive X--ray
coverage).

\begin{figure}[t!]
\begin{center} \hbox{
\psfig{figure=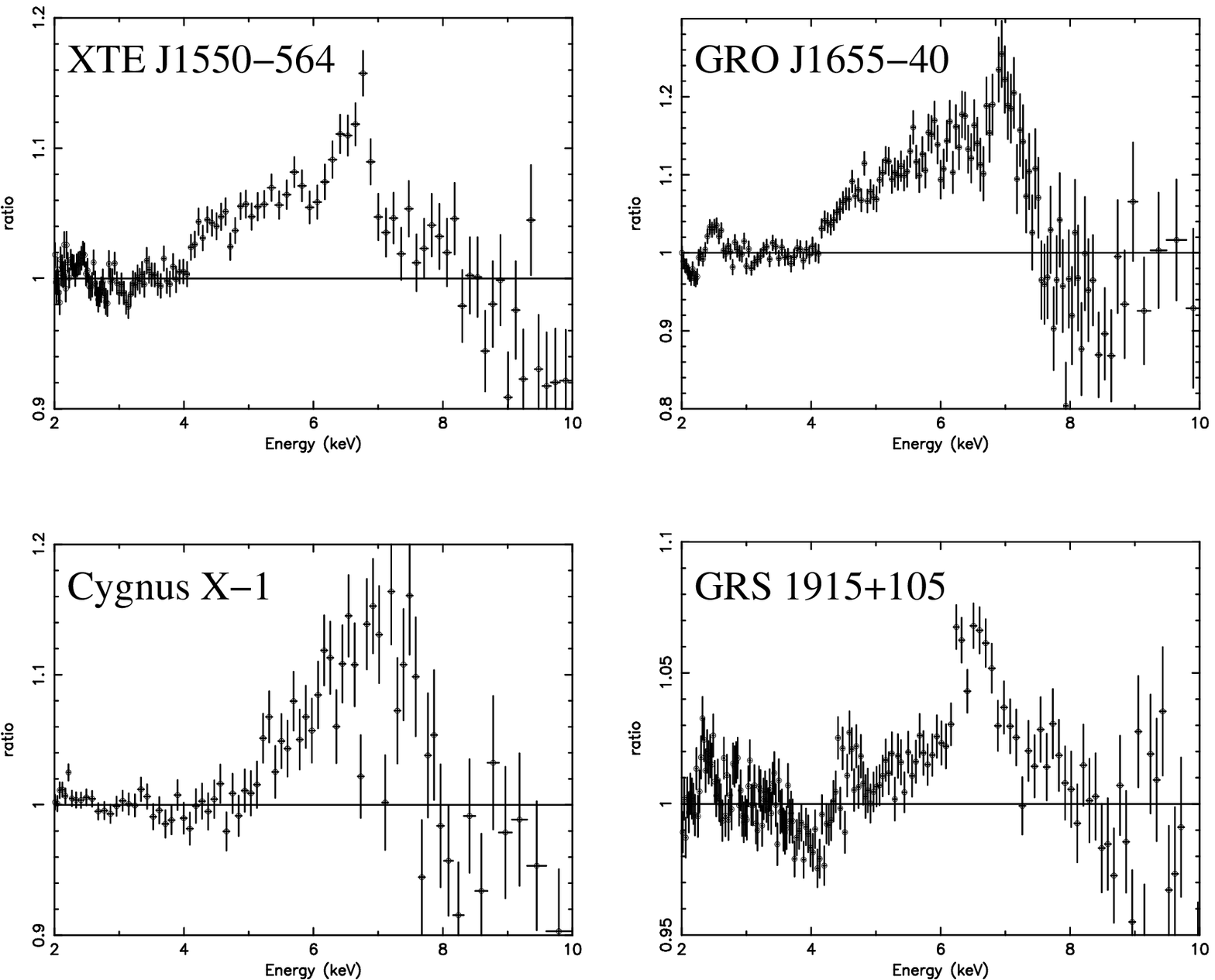,width=0.45\textwidth,height=0.32\textwidth}
\hspace{0.8cm}
\psfig{figure=gx339_xmm.ps,width=0.45\textwidth,height=0.32\textwidth,angle=-90}
\hspace{1.0cm}
}
\vspace{-1cm} 
\end{center} 
\caption{\footnotesize {{\it Left panel}: Prominent relativistic lines
    observed with \asca\ in GBHC. When fitted with relativistically
    blurred reflection models all sources but GRS~1915+105 require
    emission from within the ISCO of a non--rotating BH, suggesting
    that the BHs in most sources are rapidly spinning.  {\it Right
      panel}: The broad iron line in GX~339--4 as observed by \xmm\
(Miller et al 2004b).  }
}
\label{jonasca} 
\end{figure}

Examples of broad Fe lines detected with \asca\ in the X--ray spectra
of GBHcs are shown in the left panel of Fig.~\ref{jonasca} from Miller
et al (2004). Other broad lines have been found in other GBHCs,
especially in the high and very high/intermediate states where the
accretion disc is believed to extend down to small radii (Martocchia
et al 2002; Miller et al 2002a,b,c; 2003). In the right panel of the
same Figure, we show the \xmm\ Fe line profile of GX~339--4 (Miller et
al 2004b). A similar profile is detected by \axaf\ as well (Miller et
al 2004a).
 
\subsection{The case of XTE~J1650--500}

\begin{figure}[t!]
\begin{center} 
  \hbox{
    \psfig{figure=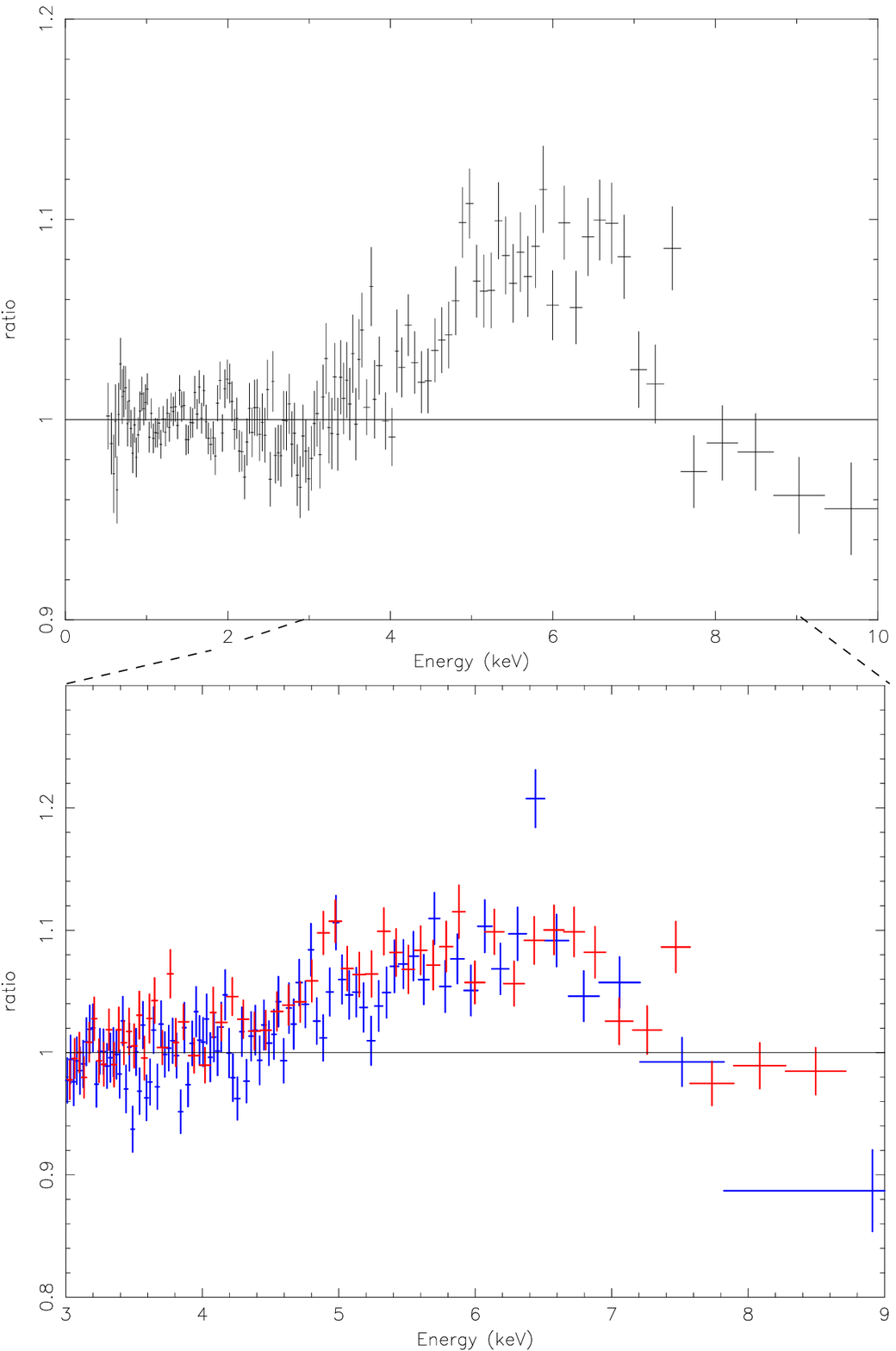,width=0.45\textwidth,height=0.32\textwidth}
    \hspace{0.8cm}
    \psfig{figure=rossi1650a.eps,width=0.45\textwidth,height=0.32\textwidth}
  } \hbox{
    \psfig{figure=gio1650.ps,width=0.45\textwidth,height=0.32\textwidth,angle=-90}
    \hspace{0.8cm}
    \psfig{figure=rossi1650b.eps,width=0.45\textwidth,height=0.32\textwidth}
  } \vspace{-1cm} \end{center} \caption{\footnotesize {{\it
      Left--top panel}: The line profile of XTE~J1650--500 as observed
    with \xmm\ (Miller et al 2002a). {\it Left--middle panel}:
    Superposition of the broad lines in XTE~J1650--500 (red) and
    Cyg~X--1 (blue) from Miller et al 2002a. {\it Left--bottom panel}:
    Broadband \sax\ spectrum of XTE~J1650--500 (as a ratio to the
    continuum). The signatures of relativistically--blurred reflection
    are clearly seen. {\it Rigt--top panel}: The broad Fe line flux is
    plotted versus the ionizing continuum. {\it Right--bottom panel}:
    The ratio between the line flux and the continuum (i.e.\ a measure
    of the reflection fraction) is plotted against the ionizing flux.
    Both right panels are from Rossi et al (2005) and refer to \xte\
    data.  }} \label{1650}
\end{figure}

XTE~J1650--500 is a X--ray binary with a period of about 7.6 hours and
an optical mass function $f(M)=2.73\pm 0.56~M_\odot$. If typical mass
ratios with the companion and an inclination of about $50^\circ$ are
assumed (the latter in agreement with X--ray spectroscopy) the mass of
the primary turns out to be $7.3\pm 0.6~M_\odot$ strongly suggesting
the presence of a central BH (Orosz et al.\ 2004). The source outburst
was followed up during 2001 by \sax\ (three observations)
\xmm\ (one observation) and \xte\ itself (57 pointed observations).
The \xmm\ observation revealed the presence of a broad relativistic
line (Miller et al 2002a) which is shown in the top--left panel of
Fig.~\ref{1650}. The line profile extends down to about 4~keV and
implies emission from within the ISCO of a non--rotating BH. An inner
disc radius at the ISCO of a Maximal Kerr hole ($\simeq 1.24~r_g$) is
prefered over one at the ISCO of a Schwarzschild BH at the 6$\sigma$
level, strongly suggesting that the BH is rapidly, possibly maximally,
spinning. The emissivity profile is much steeper than standard and
implies that energy dissipation preferentially occurs in the inner few
$r_g$ from the center.

The source was observed three times by \sax\ and in all cases a broad
relativistic Fe line was detected confirming the main results of
Miller et al (2002a) with respect to both inner disc radius and steep
emissivity (Miniutti, Fabian \& Miller 2004). The broadband coverage
provided by \sax\ (up to 200~keV) allowed us to detect the large
reflection hump around 20--30~keV which is the unambiguous sign of the
presence of a X--ray reflection spectrum, associated with the broad Fe
line, giving more strength, if necessary, to the interpretation of the
broad spectral feature as a relativistic Fe line. The broadband 
spectrum during one of the \sax\ observations is shown in the
bottom--left panel of Fig.~\ref{1650} as a ratio with a power law (and
soft thermal emission) model.  

The steep emissivity profile and small inner disc radius are very
reminiscent of the case of \mcg. In the latter case, the variability
of the Fe line was succesfully explained by the light bendind model. It
is therefore an interesting exercise to study the Fe line variability
in XTE~J1650--500 as well and to put to test the light bending model
not only in AGNs, but in GBHCs as well. The three \sax\ observations
already gave some indication in that direction (Miniutti, Fabian \&
Miller 2004) suggesting the relevance of the light bending model for
this object. Further study by Rossi et al (2005) confirmed the first
indication in a quite spectacular way. By using the 57 pointed
observation by \xte, Rossi et al studied the broad Fe line variability
in detail during the outburst. The results of their analysis is shown
in the right panels of Fig.~\ref{1650}. The Fe line flux is correlated
with the power law continuum at low fluxes and then saturates (or even
shows some marginal evidence for anti--correlation) at higher flux
levels. Its ratio with the power law continuum flux (a proxy for the
reflection fraction) is generally anti--correlated with the continuum
and saturates only at low fluxes. Those two results match very well
the predictions of the light bending model (see e.g.\
Fig.\ref{lbending}). We therefore suggest that during the 2001
outburst of XTE~J1650--500 most of the primary emission comes from a
compact corona (or the base of a jet) located within the innermost few
$r_g$ from the BH and that strong gravitational light bending is
responsible for most of the observed variability, as in \mcg.

\section{Broad-line-free sources and nature of the soft excess}
\label{sec:broadfree}

A broad relativistic Fe line is present in some but not all AGN
spectra. On the other hand \xmm\ and \axaf\ have shown that a narrow
Fe line is almost ubiquitous in the X--ray spectrum of Seyfert
galaxies (e.g.\ Page et al 2004; Yaqoob \& Padmanabhan 2004). The
narrow line is generally interpreted as due to reflection from distant
matter (at the parsec scale or so) such as the putative molecular
torus, in good agreement with unification schemes. Here we explore in
some detail some of the possible explanations for the non--ubiquitous
detection of relativistic broad lines in sources that are expected to
be radiatively efficient and, according to the standard model, should
have an accretion disc extending down to small radii and therefore a
broad Fe line as well.

\subsection{Observational limitations}

The X--ray spectrum of Seyfert 1 and NLS1 galaxies and quasars is
characterised by the presence of a narrow Fe line and by a soft
excess, i.e.\ soft excess emission with respect to the 2--10~keV band
spectral shape. The soft excess is often interpreted as thermal
emission from the accretion disc. However, this interpretation is
somewhat controversial. For example, Gierlinski \& Done (2004) have
selected a sample of 26 radio--quiet PG quasars for which
good--quality \xmm\ observations are available and tried to
characterise their soft excesses. When interpreted as thermal
emission, the soft excess in the sample has a mean temperature of
120~eV with a very small variance of 20~eV only.  On the other hand,
the maximum temperature of a standard Shakura--Sunyaev disc depends on
the BH mass and on the mass accretion rate, which can in turn be
related to the ratio between disc luminosity and Eddington luminosity
as $T_{\rm{max}} \propto M_{\rm{BH}}^{-1/4}
(L_{\rm{disc}}/L_{\rm{Edd}})^{1/4}$. For the BH masses and
luminosities of the PG quasars in the Gierlinski \& Done sample, one
would expect temperatures in the range 3--70~eV. Thus two problems
arise: i) the measured temperature of the soft excess is too uniform
in sources that exhibit large differences in BH mass and luminosity;
ii) the measured temperature is by far too high with respect to that
predicted from standard accretion disc models. The uniformity of the
soft excess spectral shape (and therefore of the inferred
temperatures) was noted before by e.g.\ Walter \& Fink (1993) and
Czerny et al (2003).

\begin{figure}[t!]
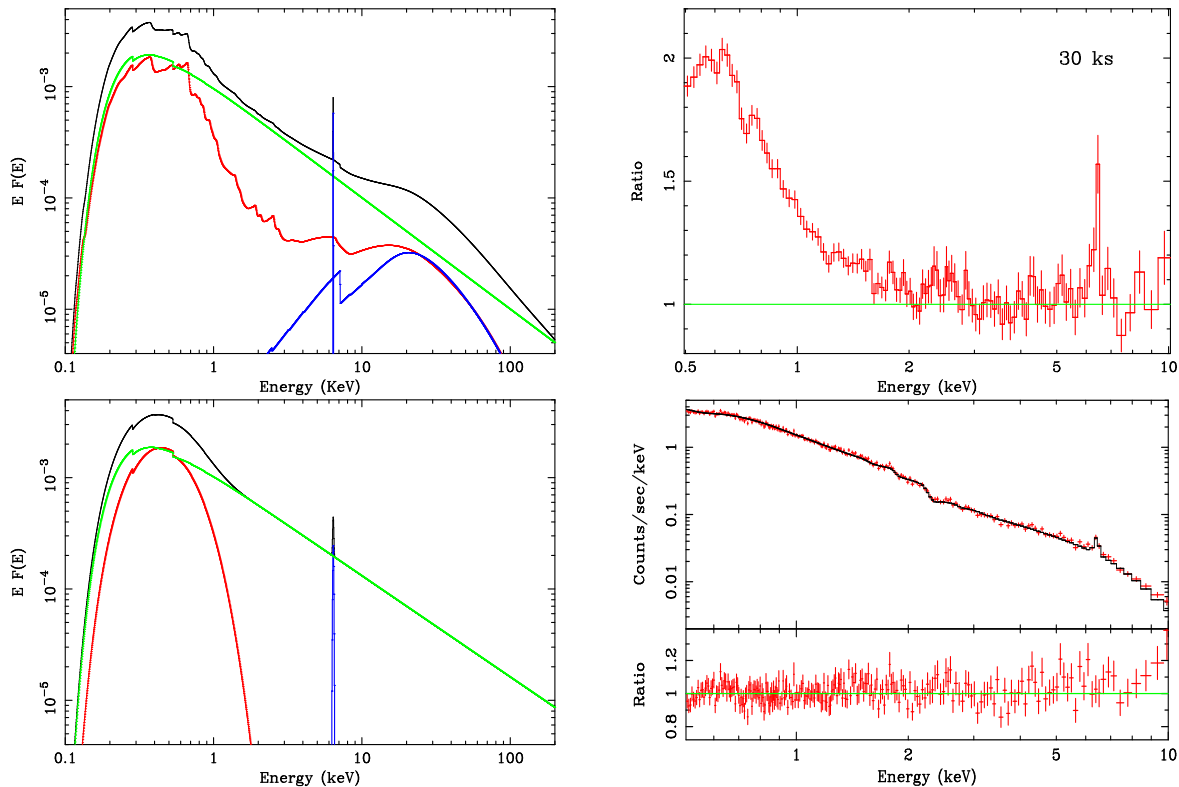
 
\begin{center} 
\hbox{
\psfig{figure=fakemo.ps,width=0.45\textwidth,height=0.32\textwidth,angle=-90}
\hspace{0.8cm}
\psfig{figure=30ks.ps,width=0.45\textwidth,height=0.32\textwidth,angle=-90}
}
%\vspace{-0.8cm} \end{center} \caption{\footnotesize {{\it Left
%panel}: 
%{\it Right panel}: }} \label{fig1} 
%\end{figure}
%\begin{figure}[] 
%\begin{center} 
\hbox{
\psfig{figure=simplemo.ps,width=0.45\textwidth,height=0.32\textwidth,angle=-90}
\hspace{0.8cm}
\psfig{figure=simple.ps,width=0.45\textwidth,height=0.32\textwidth,angle=-90}
} 
\vspace{-1cm} \end{center} \caption{\footnotesize 
{{\it Top--left panel}: The model we assumed for the simulations,
  based on our experience with X--ray data of Seyfert 1
  and NLS1 galaxies. 
{\it Top--right panel}: 30~ks \xmm\ simulated spectrum with the model
shown in the top--left panel as a ratio with a power law fitted in the
2--10~keV band. The only prominent features are a narrow 6.4~keV Fe
line and a steep soft--excess below about 1--2~keV, as generally observed.
{\it Bottom--left panel}: The ``standard model'' that is often used to fit the
X--ray spectra of Seyfert 1 and NLS1 galaxies.
{\it Bottom--right panel}: The 30~ks \xmm\ simulated spectrum is
fitted with the ``standard model'' shown in the bottom--left
panel. The fit is excellent with reduced $\chi^2$ of 0.98. 
}} \label{xmmsimu} 
\end{figure}

The uniformity and implausibly high temperature of the soft excess
raises the question on the true nature of the soft excess. A uniform
temperature could be easily understood if the soft spectral shape was
due to atomic rather than truly thermal processes, the obvious
candidate being ionized reflection from the disc. The uniformity and
too large temperature of the soft excess may be then the result of
applying the wrong spectral model (thermal emission) to the data. When
a broad Fe line is present (e.g.\  \mcg, 1H~0707--495) the parameters
of the reflection and relativistic blurring are all constrained by the
line shape and energy. It is then remarkable that the same model, once
extrapolated in the soft band, provides an excellent description of
the broadband spectrum and does not need any additional soft excess.
We therefore suggest that relativistically blurred ionized reflection
may well be responsible for the soft excess in many other objects as
well (see also Ross \& Fabian 2005; Crummy et al, in preparation).

We then construct a theoretical model for what we believe could well
be the typical X--ray spectrum of radiatively efficient X--ray
sources. The model comprises a power law component and reflection from
distant matter (represented by neutral reflection continuum and narrow
Fe line). The soft excess is provided by ionized reflection from the
accretion disc that also contributes to the hard band mainly with a
broad relativistic line. We assume solar abundances and isotropic
illumination of the accretion disc with reflection fraction $R\sim 1$.
These conditions are less favourable for the detection of the
relativistic Fe line than those found for example in \mcg\ and
1H~0707--495, in which the Fe abundance is super--solar and the RDC
contribution particularly high, and should represent the typical and
most common situation. The ionized reflection spectrum is convolved
with a Laor model reproducing the Doppler and gravitational effects in
an accretion disc around a Maximal Kerr BH (the other parameters being
the disc inclination and the emissivity index, chosen to be
$i=30^\circ$ and $q=3$).

The model is shown in the top--left panel of Fig.~\ref{xmmsimu}: a
flux of $3\times 10^{-12}$~erg~cm$^{-2}$~s$^{-1}$ in the
2--10~keV band is assumed. We then simulated an \xmm\ observation with
a typical exposure time of 30~ks. The simulated spectrum is shown in
the top--right panel of Fig.~\ref{xmmsimu} as a ratio with a power law
in the 2--10~keV band. The main features in the simulated spectrum are
a narrow 6.4~keV Fe K$\alpha$ line and a soft excess below 1--2~keV,
as generally observed in Seyfert galaxies and quasars. The
broad Fe line is completely lost into the continuum and no detection
can be claimed with any significance. In fact, the spectrum can be
described by the generally adopted model (hereafter ``standard
model'') shown in the bottom--left panel of Fig.~\ref{xmmsimu}
comprising a power law, narrow Fe line, and blackbody emission to model
the soft excess. The fit with this model is shown in the bottom--right
panel of the same figure and is excellent. It should be noted that the
measured temperature of the soft excess is $kT=125 \pm 15$~eV, exactly
in the range in which the ``uniform temperature'' of the soft excess
is found. 
\begin{figure}[]
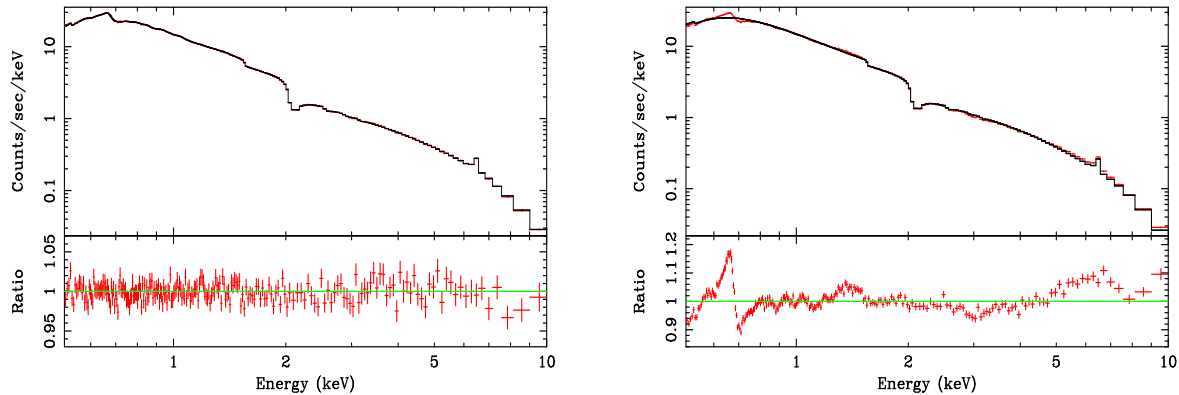
 
\begin{center} 
  \hbox{
    \psfig{figure=conxtot.ps,width=0.45\textwidth,height=0.32\textwidth,angle=-90}
    \hspace{0.8cm}
    \psfig{figure=conxsimple.ps,width=0.45\textwidth,height=0.32\textwidth,angle=-90}
  } \vspace{-1cm} \end{center} \caption{\footnotesize {{\it Left
      panel}: 100~ks Constellation--X (or XEUS with similar results)
    simulated spectrum with the same model as for the 30~ks \xmm\
    simulation. {\it Right panel}: The Constellation--X simulated
    spectrum is fitted with the ``standard model'' The fit is now
    totally unacceptable. The broad Fe line is now clearly seen in the
    4--7~keV band, and residuals corresponding to reflection features
    are left at all energy of the spectrum.  Future large collecting
    area missions such as XEUS and Constellation--X will therefore be
    able to distinguish easily between the two proposed models.}}
\label{conxsimu} 
\end{figure}

The non--detection of the broad relativistic line is also
very significant. This means that if the soft excess is due to ionized
reflection from the disc, the associated broad Fe line cannot be
detected with present--quality data unless some specific conditions
are met such as high Fe abundance, large RDC contribution, etc. This
would naturally explain not only the nature of the soft excess in many
sources but also the reason why the broad relativistic line is not
ubiquitously detected in radiatively efficient sources. In the
future, when X--ray missions with much larger effective area at 6~keV
will be launched (XEUS/Constellation--X) our interpretation can 
be tested against data. In Fig.~\ref{conxsimu} we show the same
spectrum simulated for a 100~ks Constellation--X observation. In the
left panel, we show the spectrum and the ratio with the ``real model''
(top--left panel of Fig.~\ref{xmmsimu}), while in the right panel, the
fit is made by using the ``standard model'' (bottom--left of
Fig.~\ref{xmmsimu}. The ``standard model'' clearly is an unacceptable
fit to the data and leaves residuals throughout the observed band,
most remarkably the broad Fe line between 4~keV and 7~keV as well as
soft residuals due to unmodelled soft features in the reflection
spectrum.   

\subsection{Generalization of the light bending model}

Good examples of broad--line--free sources from long \xmm\ exposures
(among others) are Akn\,120, which has no warm absorber (Vaughan et
al.\ 2004), the Seyfert 1 galaxy NGC~3783 (Reeves et al 2004), and the
broad line radio galaxy 3C\,120 (Ballantyne, Fabian \& Iwasawa 2004).
Various possibilities for the lack of any line have been proposed by
the authors of those and other papers including: a) the central part
of the disc is missing; b) the disc surface is fully ionized (i.e.\
the Fe is); c) the coronal emissivity function is flat, which could be
due to d) the primary X-ray sources being elevated well above the disc
at say $100 r_{\rm g}$.

There are also intermediate sources where the data are either poor or
there are complex absorption components so that one cannot argue
conclusively that there is a relativistic line present. Some narrow
line components are expected from outflow, warm absorbers and distant
matter in the source. One common approach in complex cases, which is
not recommended, is to continue adding absorption and emission
components to the spectral model until the reduced $\chi^2$ of the fit
is acceptable, and then claim that model as the solution. Very broad
lines are difficult to establish conclusively unless there is
something such as clear spectral variability indicating that the
power-law is free of Fe-K features, as found for MCG--6-30-15, or for
GX\,339-4 and XTE~J1650--500 where the complexities of an AGN are not
expected.

Our interpretation of the spectral behaviour of \mcg\ and some other
sources means that we are observing the effects of very strong
gravitational light bending within a few gravitational radii of a
rapidly spinning black hole. The short term (10--300~ks) behaviour is
explained, without large intrinsic luminosity variability, through
small variations in the position of the emitting region in a region
where spacetime is strongly curved.  This implies that some of the
rapid variability is due to changes in the source position. Now BHC in
the (intermediate) high/soft state have high frequency breaks at
higher frequency, for the same source, than when in the low/hard state
(cf. Cyg X-1, Uttley \& McHardy 2004).  This additional variability
when in the soft state could be identified with relativistic
light-bending effects on the power-law continuum.

This picture suggests a possible generalization of the light-bending
model to unify the AGN and GBHC in their different states.  Note the
work of Fender, Belloni \& Gallo (2004) which emphasises that jetted emission
occurs commonly in the hard state of GBHC. The key parameter may be the
mean height of the main coronal activity above the black hole. Assume
that much of the power of the inner disc passes into the corona
(Merloni \& Fabian 2001a,b) and that the coronal activity is magnetically
focused close to the central axis. Then at low Eddington ratio the
coronal height is large (say $100r_{\rm g}$ or more), the corona is
radiatively inefficient and most of the energy passes into an outflow;
basically the power flows into a jet. Reflection is then appropriate for
Euclidean geometry and a flat disc and there is only modest broadening to the
lines. If the X-ray emission from the (relativistic) jet dominates
then X-ray reflection is small (see e.g.\ Beloborodov 1999). The high
frequency break to the power spectrum is low ($\sim 0.001c/r_{\rm
g}$). 

When the Eddington fraction rises above say ten per cent, the height
of the activity drops below $\sim 20 r_{\rm g}$, the corona is more
radiatively efficient and more high frequency variability occurs due
to light bending and the turnover of the power spectrum rises above
$0.01c/r_{\rm g}$. The X-ray spectrum is dominated at low heights by
reflection, including reflection-boosted thermal disk emission, and a
broad iron line is seen. Any jet is weak. The objects with the highest
spin and highest accretion rate give the most extreme behaviour.
Observations suggest that these include NLS1 and some very high state,
and intermediate state, GBHC. Some broad-line-free sources do not
however fit this model, so more work is required also to match the
observed timing properties and not only the spectral variability.

\section{Short--timescale variability in the Fe K band}

In addition to the major line emission around 6.4 keV, transient
emission features at energies lower than 6.4 keV are sometimes
observed in X-ray spectra of AGN. As discussed, an early example was
found in the ASCA observation of MCG--6-30-15 in 1997, which was
interpreted as Fe K emission induced from a localised region of the
disc, possibly due to illumination by a flare above it (Iwasawa et al
1999, see Fig.~\ref{ki99}).  More examples followed in recent years
with improved sensitivity provided by XMM-Newton and Chandra X-ray
Observatory (Turner et al 2002; Petrucci et al 2002; Guainazzi et al
2003; Yaqoob et al 2003; Ponti et al 2004; Dov\v{c}iak et al 2004;
Turner Kraemer \& Reeves 2004; Longinotti et al 2004; Porquet et al
2004; Miniutti \& Fabian 2005).  These features can be attributed to
an Fe K$\alpha$ line arising from relatively localised reflecting
spots on the accretion disc. If the spot is close to the central black
hole, then the line emission is redshifted, depending on the location
of the spot on the disc (Iwasawa et al 1999; Ruszkowski 2000;
Nayakshin \& Kazanas 2001; Dov\v{c}iak et al 2004). Here we present
one case only, probably the most spectacular observed so far
(NGC~3516, Iwasawa, Miniutti \& Fabian 2004)

\subsection{The remarkable case of NGC~3516}

\begin{figure}[]
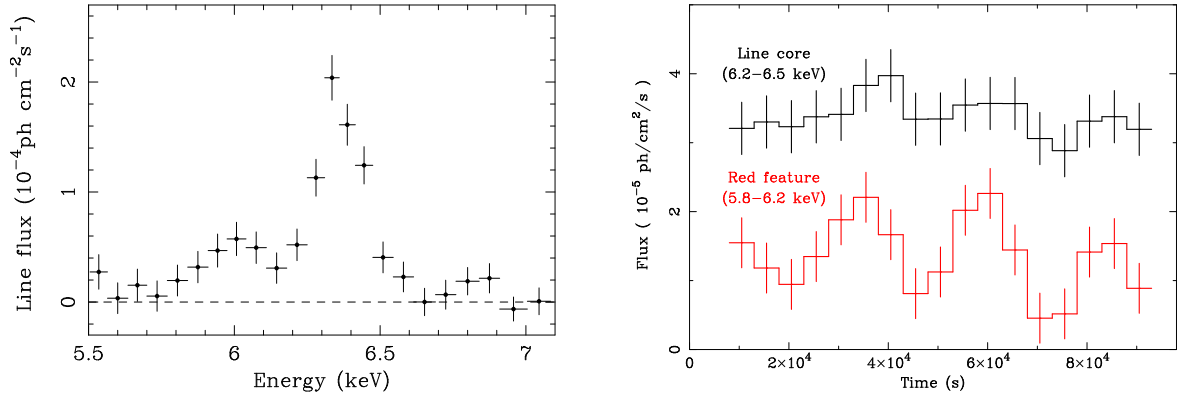
 
\begin{center} 
\hbox{
\psfig{figure=3516averaged.ps,width=0.45\textwidth,angle=-90,height=0.32\textwidth}
\hspace{0.8cm}
\psfig{figure=3516lc.ps,width=0.45\textwidth,angle=-90,height=0.32\textwidth}
} 
\vspace{-1cm} 
\end{center} 
\caption{\footnotesize {{\it Left panel}: Time--averaged line(s)
    profile of NGC~3516. Two major emission features are seen, namely
    a line core at 6.4~keV and a red feature around 6~keV
    (rest--frame).  {\it Right panel}: Light curves of the line core
    and red feature on a 5~ks timescale. The red feature seems to vary
    on a characteristic timescale of 25~ks while the line core is
    largely constant, apart from a possible increase delayed from the
    'on' phase in the red feature by a few ks.}}
\label{3516one} 
\end{figure}

Iwasawa, Miniutti \& Fabian (2004) selected one of the XMM-Newton
observations of the bright Seyfert galaxy NGC3516, for which Bianchi
et al (2004) reported excess emission at around 6~keV in addition to a
stronger 6.4 keV Fe K$\alpha$ line in the time-averaged EPIC spectrum
and studied the short--timescale variability of the 6~keV emission
feature. 
%The broad-band X-ray spectrum of NGC3516 is very complex, as
%a result of modification by absorption and reflection. Since our
%interest is in the behaviour of the relatively narrow feature in the
%Fe K band, we designed our analysis method as follows to avoid
%unnecessary complications: 1) The energy band is restricted to 5.0--7.1
%keV, which is free from absorption which can affect energies below and
%above; 2) the continuum is determined by fitting an absorbed power-law
%to the data excluding the line band, and is subtracted to obtain
%excess emission which is then corrected for the detector response.
%When determining the continuum in this way, the spectral curvature
%likely induced by a broad Fe line emission is approximated by the
%effect of absorption and its contribution subtracted away together
%with the underlying continuum. 
 We consider the 5--7.1~keV band, free from absorption which
can affect energies above and below and fit an absorbed power law
model to the data. The only residual emission features in
the time--averaged spectrum are then two relatively narrow emission
lines at 6.4~keV and 6~keV (hereafter the ``line core'' and ``red
feature'' respectively).  This is shown in the left panel of
Fig.~\ref{3516one} where the line(s) profiles detected in the
time--averaged spectrum are plotted.

\begin{figure}[]
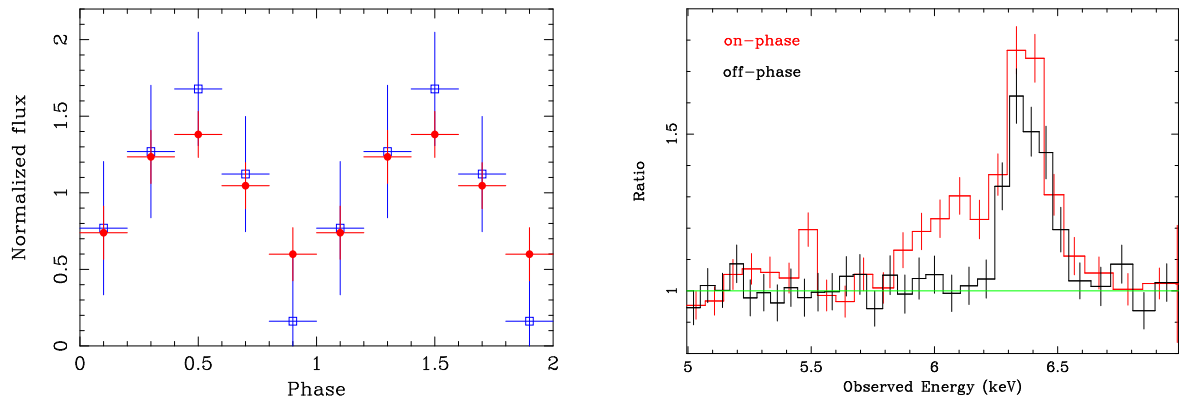
 
\begin{center} 
\hbox{
\psfig{figure=3516folded.ps,width=0.45\textwidth,angle=-90,height=0.32\textwidth}
\hspace{0.8cm}
\psfig{figure=3516lines.ps,width=0.45\textwidth,angle=-90,height=0.32\textwidth}
} 
\vspace{-1cm} 
\end{center} 
\caption{\footnotesize {{\it Left panel}: Red feature folded light
    curve on 25~ks from the smoothed (red) and unsmoothed (blue)
    images. See Iwasawa, Miniutti \& Fabian (2005) for more details.
    {\it Right panel}: The line profile (shown as a ratio to the
    continuum) for the on-- and off--phases. The red feature is
    present in the on--phase only, the difference with the off--phase
    spectrum being above the 4 $\sigma$ level.}} \label{3516two}
\end{figure}

We first investigated the excess emission at resolutions of 5 ks in
time and 100 eV in energy. A smoothed image of the excess emission in
the time-energy plane is constructed from individual intervals of
5~ks.  The detailed procedure of this method is described in Iwasawa,
Miniutti \& Fabian (2005). The light curves of the line core at 6.4
keV (6.2--6.5 keV) and of the red feature (5.8--6.2 keV) are obtained
from the image and shown in the right panel of Fig.~\ref{3516one}. The
errors on the line fluxes are estimated from extensive simulations as
discussed in Iwasawa, Miniutti \& Fabian (2005). The red feature
apparently shows a recurrent on-and-off behaviour which is suggestive
of about four cycles with a timescale of 25~ks. In contrast, the
6.4 keV line core remains largely constant, apart from a possible
increase delayed from the `on' phase in the red feature by a few ks.
Folding the light curve of the red feature confirms a
characteristic interval of about 25~ks. The left panel of
Fig.~\ref{3516two} shows the folded light curve of the red feature
(red), as well as the one obtained from the original unsmoothed data
(blue) by folding on a 25 ks interval.

Using the light curves as a guide (see right panel of
Fig.~\ref{3516one}), we constructed two spectra by selecting intervals
in a periodic manner from the on and off phases to verify the
implied variability in the red feature. The line profiles obtained
from the two spectra are shown in the right panel of
Fig.~\ref{3516two}. The 6.4 keV core is resolved slightly ($\sim 5,000$ \kmps
in FWHM) and found in both spectra with an equivalent width of
110~eV.  While the 6.4 keV core remains similar between the two, there
is a clear difference in the energy range of 5.7--6.2 keV due to the
presence/absence of the red feature. The variability detected between
the two spectra is significant at more than the $4\sigma$ level.

\begin{figure}[] 
\begin{center} 
\hbox{
\psfig{figure=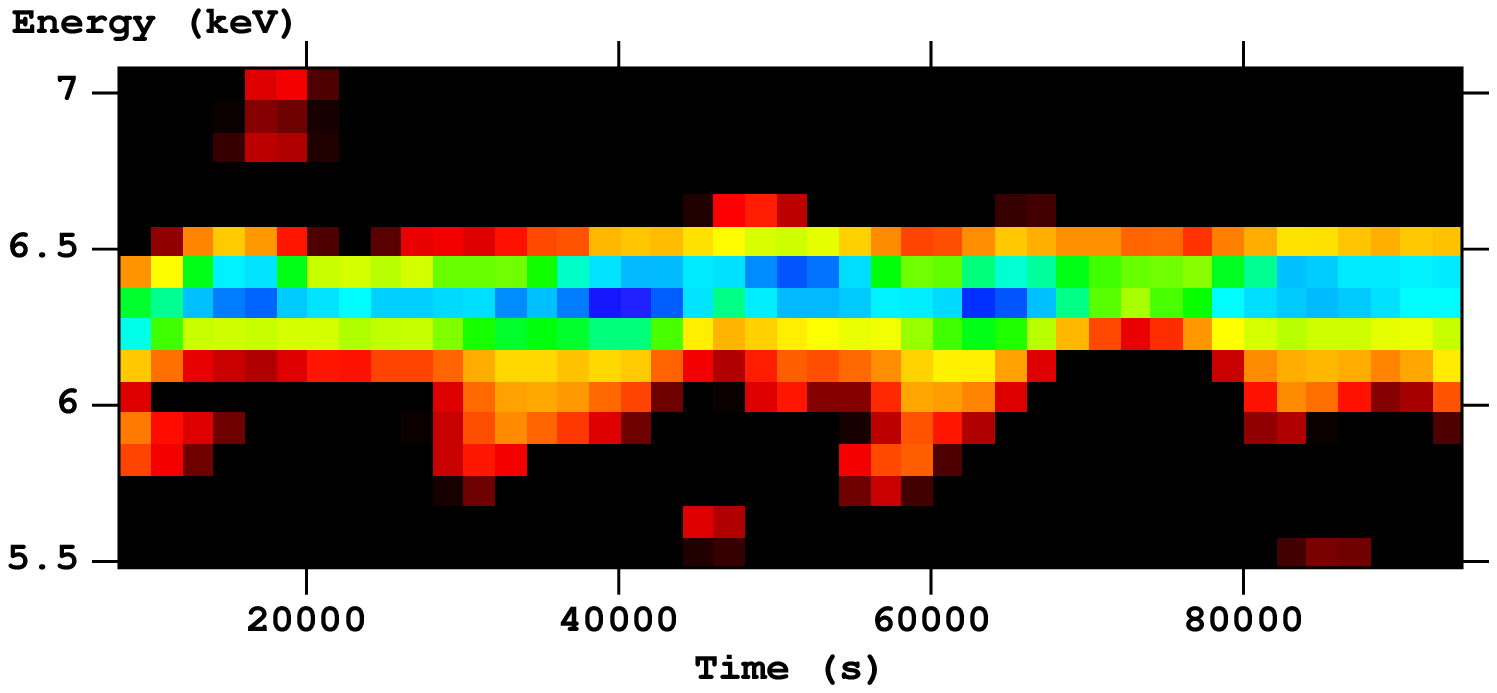,width=0.46\textwidth,height=0.3\textwidth}
\hspace{0.38cm}
\psfig{figure=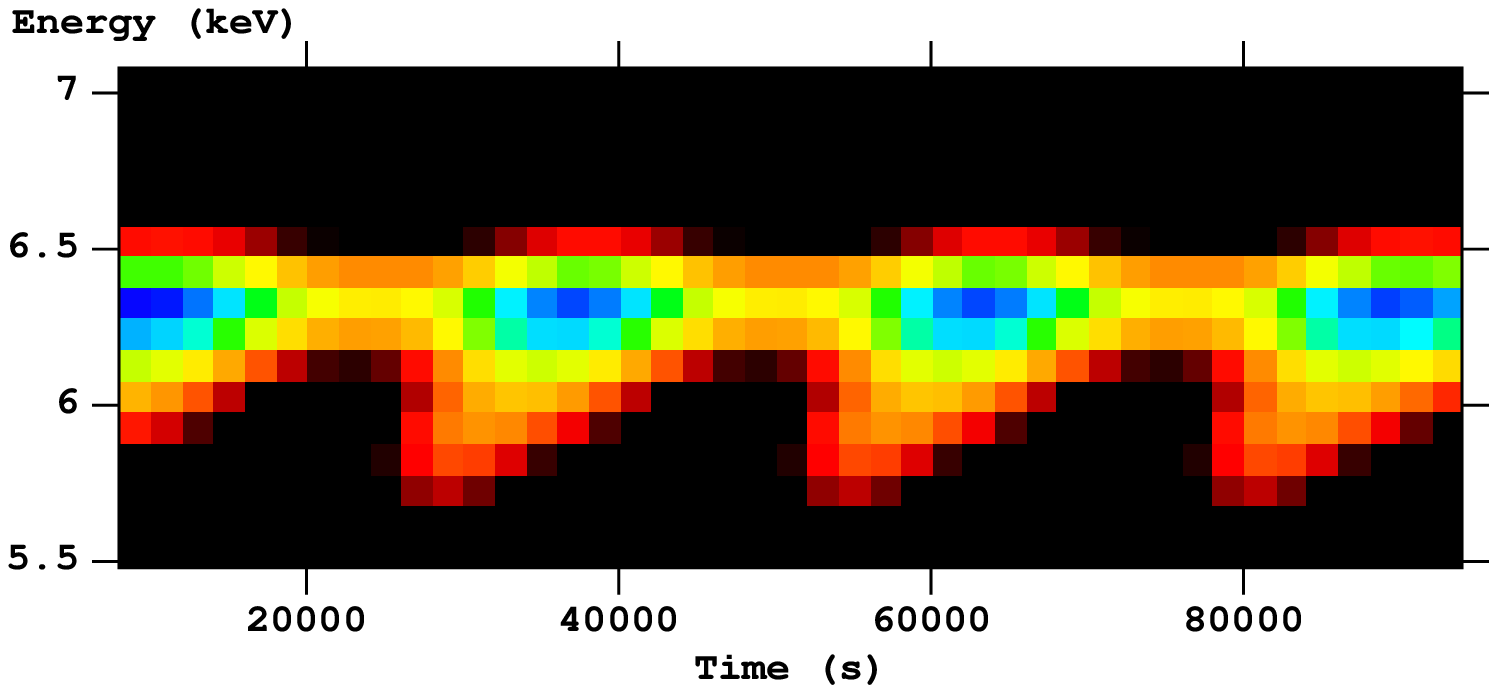,width=0.46\textwidth,height=0.3\textwidth}
} 
\vspace{-1cm} 
\end{center} 
\caption{\footnotesize {{\it Left panel}: Excess emission image in the
    time--energy plane from the \xmm\ observation. Each pixel is 2~ks
    in time and 100~eV in energy.  {\it Right panel}: Theoretical
    image in the time--energy plane. The theoretical model assumes the
    presence of a flare orbiting the BH at $9~r_g$ and at an height of
    $6~r_g$ plus a constant line core at 6.4~keV. The match with the
    observed image (left) is surprisingly good.}} \label{3516images}
\end{figure}

We then investigated the variability of the red feature on shorter
timescales to establish if not only flux, but also energy variation
can be detected, with the ultimate goal of inferring the origin of the
red feature. We constructed an image in the time--energy plane with
2~ks resolution in time and 100~eV in energy. The resulting image is
seen in the left panel of Fig.~\ref{3516images} where the colour scale
indicates the number of photons detected above the continuum (dark
blue corresponding to about 15 counts). The image shows a relatively
stable line core around 6.4~keV, while the red feature apparently
moves with time during each on phase: the feature emerges at around
5.7 keV, shifts its peak to higher energies with time, and joins the
major line component at 6.4 keV, where there is marginal evidence for
an increase of the 6.4 keV line flux. This evolution appears to be
repeated for the on--phases. 

The detection of only four cycles is not sufficient to establish any
periodicity at high significance. The 25 ks is however a natural
timescale of a black hole system with a black hole mass of a few times
$10^7~M_\odot$, as measured for NGC~3516 (e.g Onken et al 2003;
Peterson et al 2004). The finding could potentially be important and,
especially considering the evolution of the line emission, warrant a
theoretical study. We adopt a simple model in which a flare is located
above an accretion disc, corotating with it at a fixed radius.  The
flare illuminates an underlying region on the disc (or spot) which
produces a reflection spectrum, including an Fe K$\alpha$ line. The
observed line flux and energy are both phase-dependent quantities and
therefore, if the flare lasts for more than one orbital period, they
will modulate periodically. We therefore interpret the characteristic
timescale of 25~ks we measure as the orbital period.  In the right
panel of Fig.~\ref{3516images} we show a theoretical image constructed
by considering a flare orbiting at $9~r_g$ from the BH. The
theoretical model matches the data well enough. It is also possible
that the emitting spot is not due to illumination from a flare but to
some structure with enhanced emissivity (due to reflection of a more
central continuum) on the disc itself (overdensity, spiral wave, etc).

By constructing a large number of theoretical images and comparing
them with the data we could constrain the location of the reflecting
spot (or illuminating flare) in the range $7$--$16~r_g$ from the BH.
By combining this result with the timescale of 25~ks and by assuming
this is the orbital period, the mass of the central BH in NGC~3516 can
be estimated to be $1$--$5\times 10^7 M_\odot$, in excellent agreement
with results obtained through reverberation mapping of optical
emission lines. The
estimates of $M_{\rm{BH}}$ in NGC~3516 from reverberation mapping lie
in the range between $1 \times 10^7~M_\odot$ and $4 \times
10^7~M_\odot$. Onken et al (2003) derive a value of $(1.68 \pm 0.33)
\times 10^7~M_\odot$, while Peterson et al (2004) estimate $(4.3 \pm
1.5)\times 10^7~M_\odot$.  The previous analysis based on H$\beta$
alone gave $2.3\times 10^7~M_\odot$ (Ho 1999; no uncertainty is
given). It is remarkable that our estimate of the black hole mass in
NGC~3516 is in excellent agreement with the above results. Although
the systematic flux and energy variability we report here is only
tentative, the above agreement supports our interpretation.

Events like this happening at smaller radii where gravity is stronger
are at present beyond the capabilities of X--ray observatories as is
X--ray (Fe line) reverberation mapping (Reynolds et al 1999; Young \&
Reynolds 2000).  It is clear from the results presented above that the
potential of future missions with much larger collecting area than
XMM--Newton in the Fe K band, such as XEUS/Constellation--X, is
outstanding. The prospects of probing the strong gravity regime of
General Relativity via X--ray observations look stronger now than ever
before.

%\begin{figure}[] 
%\begin{center} 
%\hbox{
%\psfig{figure=1425half1.ps,width=0.45\textwidth,angle=-90,height=0.36\textwidth}
%%\psfig{figure=3516data.ps,width=0.48\textwidth,height=0.42\textwidth}
%\hspace{0.5cm}
%\psfig{figure=1114one.ps,width=0.45\textwidth,angle=-90,height=0.36\textwidth}
%%\psfig{figure=3516model.ps,width=0.48\textwidth,height=0.42\textwidth}
%} 
%\hbox{
%\psfig{figure=1425half2.ps,width=0.45\textwidth,height=0.32\textwidth,angle=-90}
%\hspace{0.5cm}
%\psfig{figure=1114two.ps,width=0.45\textwidth,height=0.32\textwidth,angle=-90}
%} 
%\vspace{-0.7cm} 
%\end{center} 
%\caption{\footnotesize {{\it Left
%panel}: 
%{\it Right panel}: }} \label{fig1} 
%\end{figure}

%\subsection{The future: XEUS/Constellation--X}

\section{Summary}

A relativistically-broadened iron line is unambiguous in the spectra
and behaviour of a few objects. The strength and breadth of reflection
features is strong evidence for gravitational light bending and
redshifts from a few $r_{\rm g}$. They indicate that a dense disc
extends close to the black hole, which must therefore be rapidly
spinning ($a/m>0.8$). Roy Kerr's solution to the Einstein equations
in vacuum appears to be the astrophysically relevant one and the key
ingredient in Galactic and extra--Galactic powerful sources of
radiation.

The potential for understanding the accretion flow close to a black
hole is enormous. Current observations are at the limit of
\xmm\ powers, which nevertheless has enabled a breakthrough in
understanding the spectral behaviour of \mcg\ and similar objects.
Similarities in the spectral and timing properties of AGN and BHC is
enabling further progress to be made. Studies in the near future with
{\it Suzaku} followed by XEUS/Constellation-X in the next decade will
continue to open up the immediate environment of accreting black
holes, within just a few gravitational radii, to detailed study.

\begin{center}
{\bf{Acknowledgements}}
\end{center}
Thanks to Kazushi Iwasawa, Jon Miller, Chris Reynolds and Simon
Vaughan for collaboration and many discussions. ACF thanks the Royal
Society and GM the PPARC for support.

%\theendnotes

%\end{thereferences}
%\end{article}

\end{document}